\documentclass[conf]{new-aiaa}
\usepackage[utf8]{inputenc}
\usepackage[T1]{fontenc}
\usepackage[table]{xcolor}
\usepackage{amsmath}
\usepackage{graphicx,subcaption}
\usepackage{longtable,tabularx}
\usepackage{boldline}
\usepackage{tikz}
\usetikzlibrary{positioning,arrows.meta,shapes.geometric}
\usepackage[ruled,vlined]{algorithm2e}

\usepackage{hyperref}
\hypersetup{
    colorlinks=true,
    linkcolor=blue,
    citecolor=blue,
    filecolor=magenta,      
    urlcolor=cyan,
}
\usepackage{cleveref}
\DeclareRobustCommand{\legendline}[1]{%
  \textcolor{#1}{\rule[0.5ex]{5ex}{0.25ex}}%
}

\title{Predictive Compressibility Transformation for Hypersonic Turbulent Boundary Layers with Cold Walls}
\author{Engin Danis\footnote{Assistant Professor, Department of Mechanical and Aerospace Engineering, AIAA Member}}
\affil{University of Missouri, Columbia, MO 65211}
\begin{document}

\maketitle

\begin{abstract}
Compressibility transformations are widely used to relate hypersonic zero-pressure-gradient turbulent boundary layers to incompressible reference states, but their assessment has largely focused on the apparent collapse of transformed mean velocity profiles, without enforcing a unique, Mach-independent representation of the mean shear. In this work, a stricter consistency condition is proposed, requiring that a single incompressible inner--outer model for the mean velocity gradient reproduce all transformed compressible profiles when expressed in terms of a transformed wall-normal coordinate. This condition implies collapse not only of the transformed mean velocity but also of semilocal eddy viscosity and turbulent kinetic energy production. Existing compressibility transformations are shown, using hypersonic direct numerical simulation data, to incur velocity errors of $1$--$25\%$ relative to the chosen incompressible inner--outer model, particularly for strongly cooled cases. A new forward compressible-to-incompressible transformation is then developed that constructs the transformed coordinate as a convex combination of semilocal and integral-type basis functions with coefficients modeled as functions of friction Mach number $M_\tau$ and wall heat transfer rate $B_q$. Casewise optimization yields consistency errors of $1$--$4\%$ across the available hypersonic direct numerical simulation database, and this performance is retained using simple multi-linear and multi-quadratic regressions in $(M_\tau,B_q)$. The forward transformation is subsequently embedded in an inverse incompressible-to-compressible transformation framework, which reconstructs the mean compressible state from freestream and wall conditions at a prescribed boundary layer thickness. The inverse solver recovers several key boundary layer parameters, velocity profiles, and skin friction distributions with good accuracy, and generally improves upon existing models for cold-wall hypersonic TBLs, thereby providing a physically constrained basis for near-wall modeling in hypersonic turbulent boundary layers with strong wall cooling.
\end{abstract}

\section*{Nomenclature}

{\renewcommand\arraystretch{1.0}
\noindent\begin{longtable*}{@{}l @{\quad=\quad} l@{}}
$B_q$ & wall heat transfer rate, $B_q \equiv  u_q/(\gamma u_\tau)$, dimensionless \\
$C_f$ & wall skin friction coefficient, $C_f\equiv\tau_w/(\tfrac{1}{2}\rho_\infty u_\infty^2)$ \\
$C_h$ & wall heat transfer coefficient, $C_h\equiv q_w/(\rho_\infty C_p u_\infty (T_r-T_w))$, dimensionless \\
$C_p$ & specific heat at constant pressure, $\mathrm{J/(kg\cdot K)}$ \\
$C_v$ & specific heat at constant volume, $\mathrm{J/(kg\cdot K)}$ \\
$M$  & Mach number, dimensionless \\
$M_\tau$ & friction Mach number, $M_\tau \equiv u_\tau/a_w$, dimensionless \\
$P$ & turbulent kinetic energy production, $\mathrm{Pa/s}$ \\
$Pr$ & molecular Prandtl number, $Pr\equiv0.71$, dimensionless \\
$R$ & specific gas constant, 287, $\mathrm{J/(kg\cdot K)}$ \\
$Re_\tau$ & Reynolds number based on friction velocity and wall viscosity,  $Re_\tau\equiv\rho_wu_\tau\delta/\mu_w$ \\
$Re_{\theta}$ & Reynolds number based on momentum thickness and freestream viscosity, $Re_{\theta}\equiv\rho_\infty u_\infty\theta/\mu_\infty$ \\
$T$  & temperature, $\mathrm{K}$ \\
$T_r$ & recovery temperature, $T_r\equiv\left(1+Pr^{1/3}(\gamma-1)M_\infty^2\right)T_\infty$, $\mathrm{K}$ \\
$U$ & incompressible velocity, $\mathrm{m/s}$ \\
$Y$ & incompressible wall-normal coordinate, $\mathrm{m}$ \\
$a$ & speed of sound, $a=\sqrt{\gamma R T}$, $\mathrm{m/s}$ \\
$h$ & specific enthalpy, $h=C_p T$, $\mathrm{J/kg}$ \\
$q_w$ & wall heat flux, $\mathrm{W/m^2}$ \\
$u$ & compressible velocity, $\mathrm{m/s}$ \\
$u_\tau$ & friction velocity, $u_\tau\equiv \sqrt{\tau_w/\rho_w}$, $\mathrm{m/s}$ \\
$u_q$ & heat flux velocity scale, $u_q \equiv -q_w/(\rho_w C_v T_w)$, $\mathrm{m/s}$ \\
$y$ & compressible wall-normal coordinate, $\mathrm{m}$ \\
$y_\tau$ & viscous length scale, $y_\tau\equiv\mu_w/\sqrt{\rho_w\tau_w}$, $\mathrm{m}$ \\
$\delta$ & boundary layer thickness based on 99\% of freestream velocity, $\mathrm{m}$ \\
$\gamma$ & specific heat ratio, 1.4, $\gamma \equiv C_p/C_v$, dimensionless \\
$\kappa$ & the von Karman constant, $\kappa\equiv0.41$, dimensionless \\
$\mu$ & molecular viscosity, $\mathrm{kg/(m\cdot s)}$ \\
$\mu_T$ & eddy viscosity, $\mathrm{kg/(m\cdot s)}$ \\
$\rho$ & density, $\mathrm{kg/m^3}$ \\
$\theta$ & momentum thickness,  $\mathrm{m}$\\
$\tau_w$ & wall shear stress, $\mathrm{Pa}$ \\
\multicolumn{2}{@{}l}{Subscripts}\\
$i$ & incompressible variables \\
$w$ & wall variables \\
$\infty$ & freestream variables \\
\multicolumn{2}{@{}l}{Superscripts}\\
$+$ & variable in inner wall units \\
$*$ & variable in semilocal units 
\end{longtable*}}

\setcounter{table}{0}

\section{Introduction}
There has been recent interest in predicting zero-pressure-gradient (ZPG) hypersonic turbulent boundary layers (TBLs) by developing models for the compressible near-wall region \cite{Griffin2021compressibility,kumar2022,hasan2024,Ying_Li_Fu_2025}. These approaches typically rely on the present understanding of the relationship between incompressible and compressible mean states. 

A widely held view in the compressible turbulence community is that the two flow states can be mapped onto each other by appropriately accounting for changes in the mean fluid properties. This view is based on the assumption that intrinsic compressibility effects are negligible \cite{hasan2023}, an assumption commonly referred to as Morkovin's hypothesis \cite{morkovin1962}. This has led to several models that map a compressible flow state onto an incompressible state. Most notable examples are the Van Driest (VD) transformation \cite{vanDriest1951}, Trettel and Larsson (TL) transformation \cite{trettel2016}, Volpiani transformation \cite{volpiani2020}, and Griffin--Fu--Moin (GFM) transformation \cite{Griffin2021}, while Hasan, Larsson, Pirozzoli, and Pe{\v c}nik (HLLP) argued that intrinsic compressibility effects are not negligible and developed a transformation that explicitly accounts for both property variations and intrinsic effects \cite{hasan2023}.

These are also known as forward transformations (compressible-to-incompressible) and often define an integral relation for velocity,
\begin{equation}
  U^+(Y^+) = \int_{0}^{Y^+} F(y')\,dy',
  \label{eq:U_from_F_y_star}
\end{equation}
and differ primarily in how the function $F$ and the wall-normal coordinate scaling $Y^+$ are defined. For example, the VD transformation sets $F(y^+)=(\rho/\rho_w)^{1/2}\,(du^+/dy^+)$ and $Y^+=y^+$. The VD transformation is known to perform well for hypersonic TBLs with adiabatic wall conditions, but its accuracy deteriorates in the presence of strong wall cooling \cite{duan2010,duan2011}.

It has been observed that $y^*$ collapses mean turbulent statistics better than $y^+$ \cite{Huang_Coleman_Bradshaw_1995,Zhang2018}, and, as a result, many recent compressibility transformations \cite{trettel2016,Griffin2021,hasan2023} use $Y^+ = y^*$. While the TL and GFM transformations aim to transform the entire boundary layer primarily by extending inner layer arguments, the HLLP transformation introduced the idea of combining separate transformations for the inner and outer regions.

There also exist forward transformations that do not rely on $y^*$ \cite{volpiani2020,danis2024transformation}. Instead, these approaches introduce a transformed wall-normal coordinate of the form
\begin{equation}
  Y^+ = \int_{0}^{y^+} G(y')\,dy',
  \label{eq:Y_from_G_y_plus}
\end{equation}
where $G$ models the effect of variable fluid properties on the wall-normal coordinate scaling. 

On the other hand, the idea of an inverse transformation is that, given a forward mapping between the compressible mean state and $(U^+,Y^+)$, one can reconstruct the compressible velocity and temperature profiles from a reference incompressible profile by coupling the transformation with a viscosity law and a temperature--velocity (TV) relation. In practice, this requires a closure for $\mu(T)$ (e.g.\ Sutherland-type laws) and a TV relation such as the one proposed in \cite{zhang2014genReAnalogy}, which together provide the mean density, viscosity, and temperature fields needed by the transformation for the ZPG TBL.

In this regard, Griffin et al. \cite{Griffin2021compressibility} reconstructed compressible mean profiles by applying an inverse form of their Griffin-Fu-Moin (GFM) transformation, which is closed with a viscosity law and the TV relation in \cite{zhang2014genReAnalogy}. Hasan et al. \cite{hasan2024} adopted a different route and derived an explicit mean shear model with inner and outer layer contributions; integration of this shear, together with the same TV relation as in \cite{Griffin2021compressibility}, yields the compressible mean fields. Recently, Ying et al. \cite{Ying_Li_Fu_2025} organized these ingredients into a general framework that couples inverse inner/outer scalings with a skin friction relation to predict compressible mean profiles and associated wall quantities. It is important to note that these methods use $Y^+=y^*$ in their underlying forward transformation. In contrast, Kumar and Larsson \cite{kumar2022} formulated an inverse transformation based on the Volpiani forward transformation in their modular framework, where $Y^+\ne y^*$ but is obtained from an integral relation like \Cref{eq:Y_from_G_y_plus}.

However, despite the apparent success of these transformations in collapsing mean velocity profiles qualitatively, the relationship between the choice of $Y^+$ and the underlying incompressible model remains underconstrained. In particular, it has been reported in \cite{danis2024transformation} that using $Y^+=y^*$ does not lead to a collapse of key inner layer quantities such as the semilocal eddy viscosity $\mu_T^*=\mu_T/\mu$ and turbulent kinetic energy (TKE) production $\tau_w^2 P^*=\mu_T (du/dy)^2$ across incompressible reference cases. These observations indicate that selecting $Y^+=y^*$ solely on the basis of improved collapse of velocity profiles does not guarantee consistency with an incompressible inner--outer model for the mean shear.

The present study addresses this ambiguity by introducing a stricter consistency requirement for the transformed wall-normal coordinate $Y^+$. First, a consistency condition is formulated that requires a single incompressible model for the mean velocity gradient, $dU^+/dY^+=F(Y^+)$, to reproduce incompressible DNS profiles when expressed in terms of the candidate coordinate $Y^+$. This condition is then used to construct a new forward (compressible-to-incompressible) transformation, in which $Y^+$ is obtained by solving an optimization problem to correct $y^*$ with additional basis functions built from property-weighted velocity and heat flux scales. Finally, the resulting forward transformation is embedded in an inverse (incompressible-to-compressible) transformation framework that, together with a viscosity law and a temperature--velocity relation, is used to predict compressible mean velocity and temperature profiles in ZPG hypersonic TBLs subjected to strong wall cooling.

The remainder of the paper is organized as follows. \Cref{sec:methods} introduces the consistency condition, develops the forward compressibility transformation, and formulates the inverse framework for reconstructing compressible mean fields from wall and freestream data. In \Cref{sec:results}, the consistency of several existing transformations is assessed, then the performance of the proposed forward model is quantified, and finally the inverse solver is evaluated against hypersonic DNS data, including predictions of $C_f$ and $C_h$. \Cref{sec:conclusion} summarizes the main findings and outlines implications for turbulence modeling and future work.

\section{Methods}\label{sec:methods}
In this section, the main focus is on the theoretical formulation.  In \Cref{sec:consistency}, a shear-based consistency requirement with respect to a fixed incompressible inner--outer model is formulated and used to define the target representation for transformed profiles. \Cref{sec:forward} then introduces a forward (compressible-to-incompressible) transformation constructed by combining property-weighted wall-normal coordinates with a simple parameterization in terms of wall-based quantities. Finally, in \Cref{sec:inverse}, these ingredients are assembled into an inverse (incompressible-to-compressible) framework that reconstructs compressible mean profiles and wall quantities from prescribed freestream and wall conditions. The calibration of these models and their performance are presented later in \Cref{sec:results}.

\subsection{Consistency condition for compressibility transformations}
\label{sec:consistency}

The starting point is the assumption that an incompressible zero-pressure-gradient turbulent boundary layer can be described by a universal inner--outer model for the mean velocity gradient in terms of the wall-normal coordinate $Y^+=Y/y_\tau$. For a given $Y^+$, this model is written as
\begin{equation}
  \frac{dU_{\text{model}}^+}{dY^+} = F(Y^+),
  \label{eq:F_incomp}
\end{equation}
where $U_{\text{model}}^+=U_{\text{model}}/u_\tau$ and $F(Y^+)$ represents a fixed inner--outer representation of the mean shear $dU^+/dY^+$ (for example, a mixing length expression in the inner layer combined with a wake contribution in the outer layer). Integrating \Cref{eq:F_incomp} gives
\begin{equation}
  U_{\text{model}}^+(Y^+) = \int_0^{Y^+} F(Y')\,dY'.
  \label{eq:U_model_from_F}
\end{equation}

A forward transformation from a compressible state $\{y,\rho,u,T,\tau_w\}$ to an incompressible state $(Y^+_{\text{transformed}},U^+_{\text{transformed}})$ is defined to be consistent if it satisfies
\begin{equation}\label{eq:U_model_eqv}
    U^+_{\text{transformed}}(Y^+_{\text{transformed}})
    \approx U^+_{\text{model}}(Y^+_{\text{transformed}})
\end{equation}
for all $Y^+_{\text{transformed}}\in[0,\delta_i^+]$. 

\Cref{eq:U_model_eqv} is referred to as a consistency condition because it does more than require a collapse of compressible mean velocity profiles onto incompressible ones. A consistent transformation must also collapse the semi-local scaled compressible eddy viscosity and TKE production onto their incompressible counterparts. 

From Danis and Durbin \cite{danis2024transformation}, $\mu_{T}^*=\mu_{T,i}^+$ and $P^*=P_i^+$, where $\mu_{T}^*={\mu_T}/{\mu}$ and $P^*={\mu P}/{\tau_w^2}$ denote the semi-local scaled compressible eddy viscosity and TKE production, respectively, while $\mu_{T,i}^+={\mu_{T,i}}/{\mu_w}$ and $P^+_i={\mu_w P_i}/{\tau_w^2}$ are their incompressible counterparts. Assuming that a suitable incompressible mean shear model $F(Y^+)$ exists, the incompressible state satisfies
\begin{equation}
    \mu_{T,i}^+ = \frac{1}{F(Y^+)} - 1,
    \qquad
    P^+_i = F(Y^+) - F(Y^+)^2.
\end{equation}
Thus, any forward transformation that satisfies \Cref{eq:U_model_eqv} necessarily collapses not only the mean velocity profiles but also $\mu_T^*$ and $P^*$ onto their incompressible counterparts, thereby providing a consistent representation of the compressible state.

In this study, following Hasan et al.\cite{hasan2023}, $F(Y^+)$ is chosen to be a combination of the Johnson and King mixing length model in the inner layer and Coles' wake function in the outer layer:
\begin{equation}\label{eq:F_inner_outer}
    F(Y^+) = \frac{1}{1+\kappa Y^+\left(1-\exp{\left(-\frac{Y^+}{A^+}\right)}\right)^2}
             +\frac{\Pi\pi}{\kappa\delta_i^+}\sin{\left(\frac{\pi Y^+}{\delta_i^+}\right)},
\end{equation}
where the von Kármán constant is $\kappa=0.41$, $A^+=17$, and the wake parameter is $\Pi=0.8\times(0.5^{3/4})$.

Despite its widespread use, the model for $F(Y^+)$ in \Cref{eq:F_inner_outer} is validated here against available incompressible direct numerical simulation (DNS) \cite{schlatter2010} and large eddy simulation (LES) \cite{schlatterLES2014} databases covering $Re_\tau=359$--$2480$. Let $Y_{\text{ref}}^+$ and $U_{\text{ref}}^+$ be the reference wall-normal coordinate and velocity profiles from DNS or LES, and compute
\begin{equation}
  U_{\text{model}}^+(Y_{\text{ref}}^+) = \int_0^{Y_{\text{ref}}^+} F(Y')\,dY',
  \label{eq:U_model_from_F_ref}
\end{equation}
and define the error measure
\begin{equation}
    \varepsilon_{\text{model}}
    =\max_{Y^+_{\text{ref}}\in[0,\delta_{{\text{ref}},i}^+]}\left|
    \frac{U_{\text{model}}^+(Y_{\text{ref}}^+)-U_{\text{ref}}^+(Y_{\text{ref}}^+)}
         {U_{\text{ref}}^+(Y_{\text{ref}}^+)}
    \right|\times100.
\end{equation}

\begin{figure}[!t]
  \centering
  \includegraphics[width=0.7\textwidth,trim={1.5cm 5cm 1.5cm 5cm},clip]{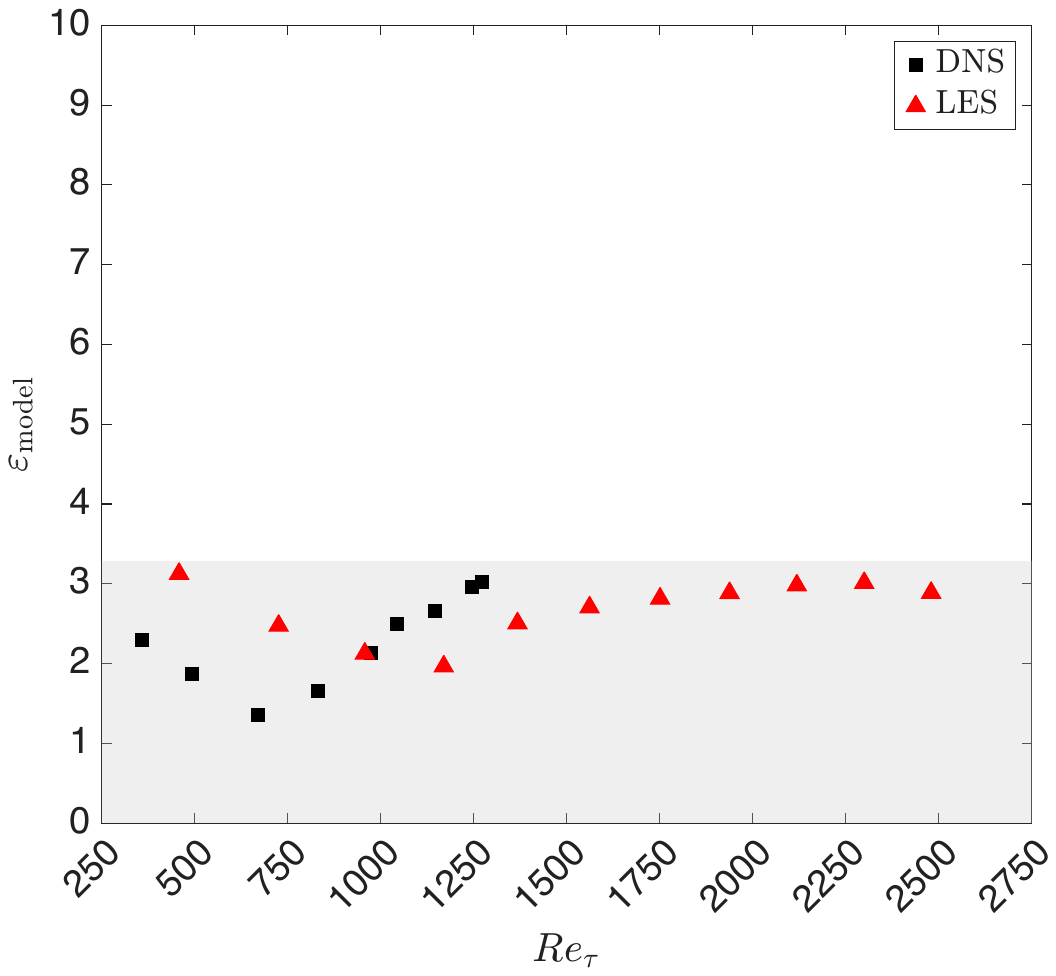}
  \caption{Maximum relative error $\varepsilon_{\text{model}}$ of the incompressible inner--outer model \Cref{eq:F_inner_outer} compared with DNS \cite{schlatter2010} and LES \cite{schlatterLES2014} reference data.}
  \label{fig:f_accuracy}
\end{figure}

\Cref{fig:f_accuracy} shows $\varepsilon_{\text{model}}$ as a function of $Re_\tau$ for the DNS and LES cases considered. The maximum error of all cases is about $3\,\%$. Therefore, \Cref{eq:F_inner_outer} is deemed sufficiently accurate to assess the consistency of forward transformations.

\subsection{Forward transformation}
\label{sec:forward}

The forward (compressible-to-incompressible) transformation maps a given compressible ZPG hypersonic TBL onto an equivalent incompressible state in terms of the inner--outer model in \Cref{sec:consistency}. For a compressible case with mean fields $u(y)$, $T(y)$, $\rho(y)$, and $\mu(y)$, the wall shear stress and wall heat flux are
\begin{equation}
  \tau_w = \mu_w \left.\frac{du}{dy}\right|_{y=0}, 
  \qquad
  q_w = -\frac{\mu_w }{Pr}\,\left.\frac{dh}{dy}\right|_{y=0},
\end{equation}
where $\mu_w = \mu(0)$, $\rho_w = \rho(0)$, and $T_w = T(0)$. The corresponding wall-based velocity scales are
\begin{equation}
  u_{\tau} = \sqrt{\frac{\tau_w}{\rho_w}},
  \qquad
  u_q = -\frac{q_w}{\rho_w C_v T_w},
\end{equation}
and a semilocal friction velocity is introduced as
\begin{equation}
  u_\tau^*(y) = \sqrt{\frac{\tau_w}{\rho(y)}}.
\end{equation}
Further parametrization is based on the friction Mach number $M_\tau$ and the wall heat transfer rate $B_q$, which are defined as
\begin{equation}
  M_{\tau} = \frac{u_\tau}{a_w},
  \qquad
  B_q = \frac{u_q}{\gamma u_\tau}.
\end{equation}

Following \Cref{sec:consistency} and Reynolds-number-based transformations in \cite{danis2024transformation}, three basis functions are defined:
\begin{align}
  Y_1^+(y) &= \frac{\rho(y)\,u_\tau^*(y)}{\mu(y)}\,y,
  \label{eq:Y1_def} \\
  Y_2^+(y) &= \int_0^{y} \frac{\rho(y')\,u_\tau^*(y')}{\mu(y')}\,dy',
  \label{eq:Y2_def} \\
  Y_3^+(y) &= \int_0^{y} \frac{\rho(y')\,u_q}{\mu(y')}\,dy'.
  \label{eq:Y3_def}
\end{align}
Note that the first basis function $Y_1^+=y^*$ is the semilocal wall-normal coordinate, while $Y_2^+$ and $Y_3^+$ incorporate the cumulative effects of variable density, viscosity, and wall cooling.  

The transformed incompressible wall-normal coordinate is then written as
\begin{equation}
  Y_{\text{inc}}^+(y) =
  A_1\,Y_1^+(y) 
  + \bigl(1-A_1\bigr)\Bigl[Y_2^+(y) - A_2\,Y_3^+(y)\Bigr],
  \label{eq:Yinc_def}
\end{equation}
where $A_1,A_2\in[0,1]$ are scalar coefficients specific to each compressible flow case. For a given $(A_1,A_2)$, \Cref{eq:Yinc_def} defines a one-to-one mapping $y \mapsto Y_{\text{inc}}^+$ as long as $dY_{\text{inc}}^+/dy>0$. This choice of \Cref{eq:Yinc_def} is based on the following observation. In the adiabatic incompressible limit, $M_\tau,B_q\rightarrow0$ as well as $\rho\rightarrow\rho_w$, $\mu\rightarrow\mu_w$, and $u_\tau^*\rightarrow u_\tau$. Thus, $Y_1^+, Y_2^+\rightarrow Y^+_{\text{inc}}$ and $Y_3^+\rightarrow0$. Therefore, for any $A_1\in[0,1]$ and $u_q\ll u_\tau^*$, \Cref{eq:Yinc_def} is a convex combination of basis functions with proper asymptotic convergence to the incompressible $Y_{\text{inc}}^+$.

The incompressible inner--outer model $F(Y^+)$ in \Cref{eq:F_inner_outer} can be decomposed into inner and outer contributions,
\begin{equation}
  F(Y^+) = F_{\text{inner}}(Y^+) + F_{\text{outer}}(Y^+;\delta_i^+),
\end{equation}
where $F_{\text{outer}}$ is the wake term associated with Coles' function and $\delta_i^+$ denotes the incompressible boundary layer edge location. For convenience \Cref{eq:U_model_from_F} is repeated below for the proposed wall-normal coordinate transformation $Y^+_{\text{inc}}$:
\begin{equation}
  U_{\text{model}}^+(Y^+_{\text{inc}}) = \int_0^{Y^+_{\text{inc}}} F(Y')\,dY'.
  \label{eq:U_model_from_F_rep}
\end{equation}

The forward transformation for the mean velocity gradient is then defined as a combination of compressible inner and outer layer contributions:
\begin{equation}
  \frac{dU_{\text{transformed}}^+}{dY_{\text{inc}}^+}
  = \frac{\mu}{\tau_w}\,\frac{du}{dy}
  + C_w\,F_{\text{outer}}\!\left(Y_{\text{inc}}^+;\delta_i^+\right),
  \label{eq:dU_trans_def}
\end{equation}
where $C_w\in[0,1]$ is a wake coefficient to model the deviation from the incompressible outer contribution due to compressibility effects. The transformed velocity profile is obtained by integration,
\begin{equation}
  U_{\text{transformed}}^+(Y_{\text{inc}}^+)
  = \int_0^{Y_{\text{inc}}^+} 
    \left[
      \frac{\mu}{\tau_w}\,\frac{du}{dy}
      + C_w\,F_{\text{outer}}(Y';\delta_i^+)
    \right]\,dY'.
  \label{eq:U_trans_from_dU}
\end{equation}

For each compressible case, the coefficients $A_1$, $A_2$, and $C_w$ are determined by minimizing the discrepancy between the transformed profile \Cref{eq:U_trans_from_dU} and the incompressible model \Cref{eq:U_model_from_F_rep} in a least squares sense,
\begin{equation}
  \min_{A_1,A_2,C_w}
  \int_0^{\delta_i^+}
    \left[
      U_{\text{transformed}}^+(Y_{\text{inc}}^+)
      - U_{\text{model}}^+(Y_{\text{inc}}^+)
    \right]^2 dY_{\text{inc}}^+,
  \label{eq:forward_opt_problem}
\end{equation}
subject to the monotonicity condition
\begin{equation}
  \frac{dY_{\text{inc}}^+}{dy} > 0
  \quad \text{for all } y\in[0,\delta].
  \label{eq:Yinc_monotone}
\end{equation}
In the numerical implementation, an unconstrained parameter vector $\boldsymbol{p}=(p_1,p_2,p_3)^T$ is mapped to the bounded coefficients through
\begin{equation}
  A_1 = \tfrac{1}{2}\bigl(1+\tanh p_1\bigr), \quad
  A_2 = \tfrac{1}{2}\bigl(1+\tanh p_2\bigr), \quad
  C_w = \tfrac{1}{2}\bigl(1+\tanh p_3\bigr),
\end{equation}
so that $0<A_1,A_2,C_w<1$ during the casewise optimization.

To enable prediction for new flow conditions without re-optimizing $\{A_1,A_2,C_w\}$, the casewise optimal values are correlated with the wall-based parameters $M_{\tau}$ and $B_q$. Two regression families are considered.

\emph{Multi-linear model:}
\begin{equation}
  \boldsymbol{\phi}_1 =
  \begin{bmatrix}
    M_{\tau} \\
    B_q
  \end{bmatrix},
  \label{eq:phi1_features}
\end{equation}

\emph{Multi-quadratic model:}
\begin{equation}
  \boldsymbol{\phi}_2 =
  \begin{bmatrix}
    M_{\tau} \\
    B_q        \\
    M_{\tau}^2 \\
    M_{\tau} B_q \\
    B_q^2
  \end{bmatrix}.
  \label{eq:phi2_features}
\end{equation}

It is worth noting that no constant terms are included in the feature vectors, so that the correlations naturally recover the incompressible adiabatic limit $\boldsymbol{\phi}\to\boldsymbol{0}$ as $M_\tau,B_q\rightarrow0$ using a generic notation $\boldsymbol{\phi}$, which is taken as either $\boldsymbol{\phi}_1$ or $\boldsymbol{\phi}_2$ depending on whether the multi-linear or multi-quadratic model is used.

The following regression forms are adopted:
\begin{align}
  A_1(M_{\tau},B_q) &= \exp\!\left(\boldsymbol{\phi}^\top \boldsymbol{\theta}_{A_1}\right),
  \label{eq:A1_reg} \\
  A_2(M_{\tau},B_q) &= \exp\!\left(\boldsymbol{\phi}^\top \boldsymbol{\theta}_{A_2}\right),
  \label{eq:A2_reg} \\
  C_w(M_{\tau},B_q) &= \boldsymbol{\phi}^\top \boldsymbol{\theta}_{C_w},
  \label{eq:Cw_reg}
\end{align}
where $\boldsymbol{\theta}_{A_1}$, $\boldsymbol{\theta}_{A_2}$, and $\boldsymbol{\theta}_{C_w}$ are constant coefficient vectors obtained by least squares fitting to a selected subset of compressible training cases. The exponential forms in \Cref{eq:A1_reg,eq:A2_reg} enforce positivity of $A_1$ and $A_2$, while the linear model in \Cref{eq:Cw_reg} is observed to provide robustness for $C_w$ during the inverse transformation stage. 

Together, \Cref{eq:Y1_def,eq:Y2_def,eq:Y3_def,eq:Yinc_def,eq:dU_trans_def,eq:U_trans_from_dU,eq:A1_reg,eq:A2_reg,eq:Cw_reg} define the proposed forward transformation used as the basis for the inverse procedure in \Cref{sec:inverse}.

\subsection{Inverse transformation}
\label{sec:inverse}

The inverse (incompressible-to-compressible) transformation reconstructs a compressible ZPG hypersonic TBL from specified freestream and wall conditions by enforcing consistency with the incompressible inner--outer model in \Cref{sec:consistency} and the forward mapping in \Cref{sec:forward}. The inputs are the freestream state $(\rho_\infty,u_\infty,T_\infty)$, the wall temperature $T_w$, the boundary layer thickness $\delta$, the gas and transport properties $(\gamma,R,Pr)$ together with a viscosity law $\mu(T)$, a temperature--velocity relation, and the regression coefficients $\boldsymbol{\theta}_{A_1}$, $\boldsymbol{\theta}_{A_2}$, and $\boldsymbol{\theta}_{C_w}$ entering \Cref{eq:A1_reg,eq:A2_reg,eq:Cw_reg}. No \emph{a priori} information on $Re_\tau$ or $Re_\theta$ is required; these quantities are obtained \emph{a posteriori} from the reconstructed profiles. In this sense, once the freestream and wall boundary conditions of a high fidelity simulation are specified, the proposed inverse solver can be used to construct the mean BL growth by varying the boundary layer thickness $\delta$, and thereby generate the corresponding $C_f$ and $C_h$ distributions.

The initial mean velocity is taken as a linear profile between the wall and the freestream,
\begin{equation}
  u^{(0)}(y) = u_\infty \frac{y}{\delta}, \qquad 0 \le y \le \delta,
\end{equation}
and the local temperature is then calculated using the model proposed in \cite{zhang2014genReAnalogy}, as also used in \cite{Griffin2021compressibility,hasan2024,Ying_Li_Fu_2025}:
\begin{equation}\label{eq:zhang-temperature}
    \frac{T}{T_w}=1 + \frac{T_{r}-T_w}{T_w}\left[\left(1-s\,Pr\right)\left(\frac{u}{u_\infty}\right)^2 + s\,Pr\left(\frac{u}{u_\infty}\right)\right]+ \frac{T_\infty-T_{r}}{T_w}\left(\frac{u}{u_\infty}\right)^2,
\end{equation}
where $s\,Pr=0.8$. The dynamic viscosity $\mu$ is obtained from Sutherland’s law, and at a given wall-normal location, the density is approximated by the ideal gas relation at constant pressure,
\begin{equation}
  \rho(y) = \rho_\infty \frac{T_\infty}{T(y)}.
\end{equation}
All numerical wall-normal derivatives and integrals are calculated using second-order finite difference approximations and the trapezoidal rule, respectively.

The inverse problem is formulated as a scalar nonlinear equation for the wall shear stress,
\begin{equation}
  f(\tau_w) = u(\delta;\tau_w) - 0.99u_\infty = 0,
  \label{eq:nonlinear_tau}
\end{equation}
where $u(y;\tau_w)$ denotes the mean velocity reconstructed for a given trial value of $\tau_w$. Therefore, when the nonlinear solver converges, the reconstructed edge velocity reaches $99 \%$ of the freestream value $u_\infty$, and the corresponding velocity profile represents the physical mean profile at the converged $\tau_w$ for the prescribed $\delta$ and the specified freestream and wall boundary conditions.

To improve robustness, the Newton iteration is carried out in terms of the logarithm of the wall shear stress,
\begin{equation}
  \sigma = \log \tau_w,
\end{equation}
and the Newton update is written for the composite residual
\begin{equation}
  g(\sigma) \equiv f\!\left(e^{\sigma}\right),
\end{equation}
so that
\begin{equation}\label{eq:newton_iteration}
    \sigma^{k+1}=\sigma^{k} - \frac{g(\sigma^k)}{g'(\sigma^k)}.
\end{equation}
at any Newton iteration $k$.

The derivative $g'(\sigma)$ is obtained by the complex perturbation method \cite{squire1998}. At each Newton iteration, the current real part of $\sigma$ is perturbed as
\begin{equation}
  \sigma \;\to\; \sigma + i \epsilon,
\end{equation}
with a small step size $\epsilon=10^{-12}$, and the corresponding complex wall shear stress is recalculated as
\begin{equation}
  \tau_w = e^{\sigma},
\end{equation}
which guarantees that the real part of $\tau_w$ remains positive, $\text{Re}(\tau_w)>0$. For this complex-valued $\tau_w$, the inverse transformation described below is applied to obtain a complex velocity profile $u(y;\tau_w)$ and the complex residual
\begin{equation}
  \tilde{f} = u(\delta;\tau_w) - 0.99u_\infty.
\end{equation}
The Newton function value and its derivative are then approximated as
\begin{equation}
  g(\sigma) \approx \text{Re}\{\tilde{f}\}, 
  \qquad
  g'(\sigma) \approx \frac{\text{Im}\{\tilde{f}\}}{\epsilon},
\end{equation}
where $\text{Re}\{\cdot\}$ and $\text{Im}\{\cdot\}$ denote the real and imaginary parts, respectively. A relaxation factor $\lambda\in(0,1]$ is used in practice,
\begin{equation}
  \sigma^{k+1} = \sigma^{k} - \lambda\,\frac{g(\sigma^k)}{g'(\sigma^k)},
\end{equation}
to improve robustness.

For a given $\tau_w$ and the current density profile, the velocity scales $u_\tau$, $u_\tau^*(y)$, and $u_q$ along with $M_\tau$ and $B_q$ are updated as in \Cref{sec:forward}. The coefficients $A_1$, $A_2$, and $C_w$ are then obtained from the regression relations \Cref{eq:A1_reg,eq:A2_reg,eq:Cw_reg}. Depending on whether the multi-linear or multi-quadratic model is selected, either $\boldsymbol{\phi}_1$ in \Cref{eq:phi1_features} or $\boldsymbol{\phi}_2$ in \Cref{eq:phi2_features} is used in \Cref{eq:A1_reg,eq:A2_reg,eq:Cw_reg}. In the implementation, a simple relaxation is applied to these coefficients, by forming a convex combination of the newly computed and previous values of $A_1$, $A_2$, and $C_w$.

With $\tau_w$, $u_\tau^*(y)$, and $u_q$ known, the basis functions $Y_1^+(y)$, $Y_2^+(y)$, and $Y_3^+(y)$ are constructed according to \Cref{eq:Y1_def,eq:Y2_def,eq:Y3_def}, and the transformed wall-normal coordinate $Y_{\text{inc}}^+(y)$ is obtained from \Cref{eq:Yinc_def}. The incompressible inner--outer model $F(Y^+)$ in \Cref{eq:F_inner_outer} is then evaluated at $Y_{\text{inc}}^+(y)$, and decomposed into inner and outer contributions,
\begin{equation}
  F(Y_{\text{inc}}^+) = F_{\text{inner}}(Y_{\text{inc}}^+) + F_{\text{outer}}(Y_{\text{inc}}^+;\delta_i^+).
\end{equation}
Imposing that the compressible mean shear profile be consistent with the incompressible model leads to
\begin{equation}
  \frac{\tau_w}{\mu(y)}\,\frac{du}{dy}
  = F(Y_{\text{inc}}^+)
    - C_w\,F_{\text{outer}}\!\left(Y_{\text{inc}}^+;\delta_i^+\right)
  = F_{\text{inner}}(Y_{\text{inc}}^+)
    + \bigl(1- C_w\bigr)\,F_{\text{outer}}\!\left(Y_{\text{inc}}^+;\delta_i^+\right),
  \label{eq:duplus_inverse}
\end{equation}
which is integrated to compute $u(y)$. This corresponds to enforcing equality between the right-hand sides of the forward gradient definition in \Cref{eq:dU_trans_def} and the incompressible model in \Cref{eq:F_incomp}. Note that $C_w=0$ implies that the incompressible outer layer model is used unchanged in the compressible transformation, whereas $C_w=1$ removes the outer layer contribution entirely.

After each Newton update of $\sigma$, the temperature $T(y)$ is recomputed from the updated velocity $u(y)$ using \Cref{eq:zhang-temperature}, and $\rho(y)$, $\mu(y)$, and $q_w$ are updated accordingly. The iteration proceeds until a combined convergence criterion based on the changes in $u(y)$, $\tau_w$, and the coefficients $A_1$, $A_2$, and $C_w$ falls below a prescribed tolerance.

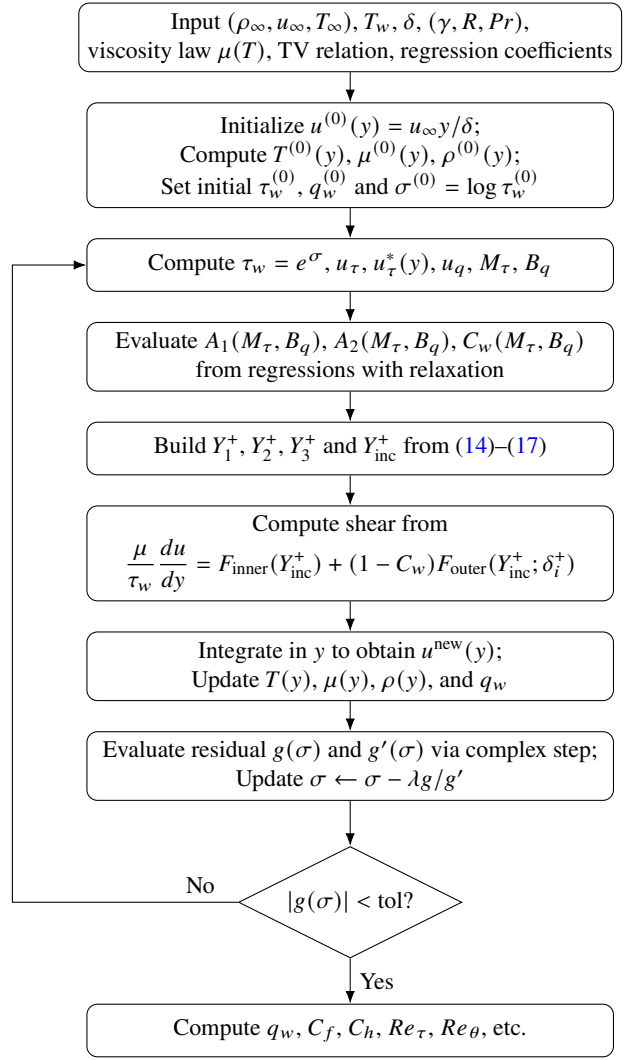
\begin{figure}[htbp]
  \centering
  \begin{tikzpicture}[
    node distance=4mm,
    every node/.style={font=\small},
    block/.style={rectangle, draw, rounded corners, align=center, minimum width=7cm, minimum height=7mm},
    decision/.style={diamond, draw, aspect=2, align=center},
    >=Latex
  ]

  \node[block] (start) {Input $(\rho_\infty,u_\infty,T_\infty)$, $T_w$, $\delta$, $(\gamma,R,Pr)$,\\
                        viscosity law $\mu(T)$, TV relation, regression coefficients};
  \node[block, below=of start] (init) {Initialize $u^{(0)}(y)=u_\infty y/\delta$;\\
                                       Compute $T^{(0)}(y)$, $\mu^{(0)}(y)$, $\rho^{(0)}(y)$;\\
                                       Set initial $\tau_w^{(0)}$, $q_w^{(0)}$ and $\sigma^{(0)}=\log\tau_w^{(0)}$};
  \node[block, below=of init] (scales) {Compute
                                        $\tau_w=e^{\sigma}$, $u_\tau$, $u_\tau^*(y)$, $u_q$, $M_\tau$, $B_q$};
  \node[block, below=of scales] (coeffs) {Evaluate $A_1(M_{\tau},B_q)$, $A_2(M_{\tau},B_q)$, $C_w(M_{\tau},B_q)$ \\ from regressions with relaxation};
  \node[block, below=of coeffs] (Yinc) {Build $Y_1^+$, $Y_2^+$, $Y_3^+$ and $Y_{\text{inc}}^+$ from \eqref{eq:Y1_def}--\eqref{eq:Yinc_def}};
  \node[block, below=of Yinc] (shear) {Compute shear from\\
    $\displaystyle \frac{\mu}{\tau_w}\frac{du}{dy}
      = F_{\text{inner}}(Y_{\text{inc}}^+)
        + (1-C_w)F_{\text{outer}}(Y_{\text{inc}}^+;\delta_i^+)$};
  \node[block, below=of shear] (uupdate) {Integrate in $y$ to obtain $u^{\text{new}}(y)$;\\
                                          Update $T(y)$, $\mu(y)$, $\rho(y)$, and $q_w$};
  \node[block, below=of uupdate] (resid) {Evaluate residual $g(\sigma)$
    and $g'(\sigma)$ via complex step;\\
    Update $\sigma \leftarrow \sigma - \lambda g/g'$};
  \node[decision, below=of resid, yshift=-2mm] (conv) {$\left|g(\sigma)\right| < \text{tol}$?};
  \node[block, below=of conv, yshift=-2mm] (post) {Compute $q_w$, $C_f$, $C_h$, $Re_\tau$, $Re_\theta$, etc.};

  \draw[->] (start) -- (init);
  \draw[->] (init) -- (scales);
  \draw[->] (scales) -- (coeffs);
  \draw[->] (coeffs) -- (Yinc);
  \draw[->] (Yinc) -- (shear);
  \draw[->] (shear) -- (uupdate);
  \draw[->] (uupdate) -- (resid);

  \draw[->] (resid) -- (conv);
  \draw[->] (conv.south) -- node[right]{Yes} (post.north);

  \draw[->] (conv.west) |- node[near start, above, xshift=-0.5cm]{No} ++(-3,0) |- (scales.west);

  \end{tikzpicture}
  \caption{Workflow of the inverse (incompressible-to-compressible) transformation.}
  \label{fig:workflow_inverse}
\end{figure}

Once convergence is achieved, all compressible mean quantities are available, including the wall shear stress and heat flux, from which the skin friction and heat transfer coefficients $C_f$ and $C_h$, the friction Reynolds number $Re_\tau$, and the momentum thickness Reynolds number $Re_\theta$ are obtained. The procedure therefore delivers a fully consistent compressible mean state that is, by construction, compatible with the incompressible inner--outer model in \Cref{sec:consistency} and the forward transformation in \Cref{sec:forward}. The overall procedure for the inverse transformation is summarized in \Cref{fig:workflow_inverse}.

\section{Results}\label{sec:results}
In this section, the consistency condition and transformation framework introduced in \Cref{sec:methods} are assessed and applied to hypersonic ZPG turbulent boundary layers. First, the consistency of several existing compressibility transformations with the incompressible inner--outer model is examined in \Cref{sec:results_consistency}. Next, \Cref{sec:results_forward} evaluates the proposed forward (compressible-to-incompressible) transformation, both in a casewise sense and in its regressed form as a function of $(M_\tau,B_q)$. Finally, \Cref{sec:results_inverse} analyzes the performance of the inverse (incompressible-to-compressible) solver in reconstructing mean velocity and temperature profiles, as well as predicting $C_f$, $C_h$, and integral boundary-layer parameters over a range of flow conditions.

\subsection{Consistency of existing compressibility transformations}
\label{sec:results_consistency}

The consistency condition in \Cref{sec:consistency} requires that the forward transformation produce a velocity profile $U_{\text{transformed}}^+(Y^+)$ that closely matches the incompressible inner--outer model $U_{\text{model}}^+(Y^+)$ obtained from \Cref{eq:U_model_from_F,eq:F_inner_outer} for the same transformed coordinate $Y^+$:
\[
  U^+_{\text{transformed}}(Y^+) \approx U^+_{\text{model}}(Y^+).
\]

In this regard, the performance of some existing compressibility transformations, namely the GFM \cite{Griffin2021}, Volpiani \cite{volpiani2020}, and HLLP \cite{hasan2023} transformations, is first assessed under this stricter requirement. These transformations are applied to the compressible DNS database of Zhang et al.\ \cite{Zhang2018} for five hypersonic ZPG TBL cases M2p5, M6Tw025, M6Tw076, M8Tw048, and M14Tw018 (see \cite{Zhang2018} for the naming convention).

In all cases, the same incompressible inner--outer model $F(Y^+)$ in \Cref{eq:F_inner_outer} is evaluated at the corresponding $Y^+$ and integrated to obtain $U_{\text{model}}^+(Y^+)$. Thus, the only difference between $U_{\text{transformed}}^+$ and $U_{\text{model}}^+$ is due to the choice of the transformation itself, where $Y^+$ is the underlying transformed wall-normal coordinate of that particular transformation. For example, $Y^+=y^*$ for the GFM and HLLP transformations, while the Volpiani transformation uses
\begin{equation}
    Y^+ = \int_{0}^{y^+}
      \left(\frac{\rho}{\rho_w}\right)^{1/2}
      \left(\frac{\mu_w}{\mu}\right)^{3/2}\,dy^+.
\end{equation}

The deviation from the incompressible model is quantified using the maximum relative error over the entire boundary layer,
\begin{equation}
    \varepsilon_{\text{transformed}}
    =\max_{Y^+\in[0,\delta_{i}^+]}\left|
    \frac{U_{\text{model}}^+(Y^+)-U_{\text{transformed}}^+(Y^+)}
         {U_{\text{model}}^+(Y^+)}
    \right|\times100,
\end{equation}
consistent with the definition in \Cref{sec:consistency}. 

\Cref{tab:consistency_errors} summarizes the maximum relative errors over the five cases considered. The maximum errors lie in the ranges $\varepsilon_{\text{GFM}} \approx 3.3$--$13.4\%$, $\varepsilon_{\text{Volpiani}} \approx 1.4$--$10.8\%$, and $\varepsilon_{\text{HLLP}} \approx 1.8$--$25.6\%$. The smallest errors occur for the low-Mach, adiabatic case (M2p5), whereas the largest discrepancies are observed for the strongly cooled hypersonic cases (M6Tw025 and M14Tw018), particularly for the HLLP model.

\begin{table}[htbp]
  \centering
  \caption{Maximum relative error in transformed velocity profiles for different compressibility transformations.}
  \label{tab:consistency_errors}
  \begin{tabular}{lccc}
    \hlineB{3}
    Case      & GFM [\%] & Volpiani [\%] & HLLP [\%] \\
    \hline
    M2p5      &  3.300 &  1.379 &  1.756 \\
    M6Tw025   & 13.391 & 10.823 &  6.273 \\
    M6Tw076   &  4.307 &  5.338 & 13.882 \\
    M8Tw048   &  4.747 &  5.176 & 16.734 \\
    M14Tw018  & 11.667 &  5.128 & 25.574 \\
    \hlineB{3}
  \end{tabular}
\end{table}

\begin{figure}[htbp]
  \centering

  \begin{subfigure}[b]{0.49\textwidth}
    \centering
    \includegraphics[width=\textwidth,trim={1.5cm 5cm 1.5cm 5cm},clip]{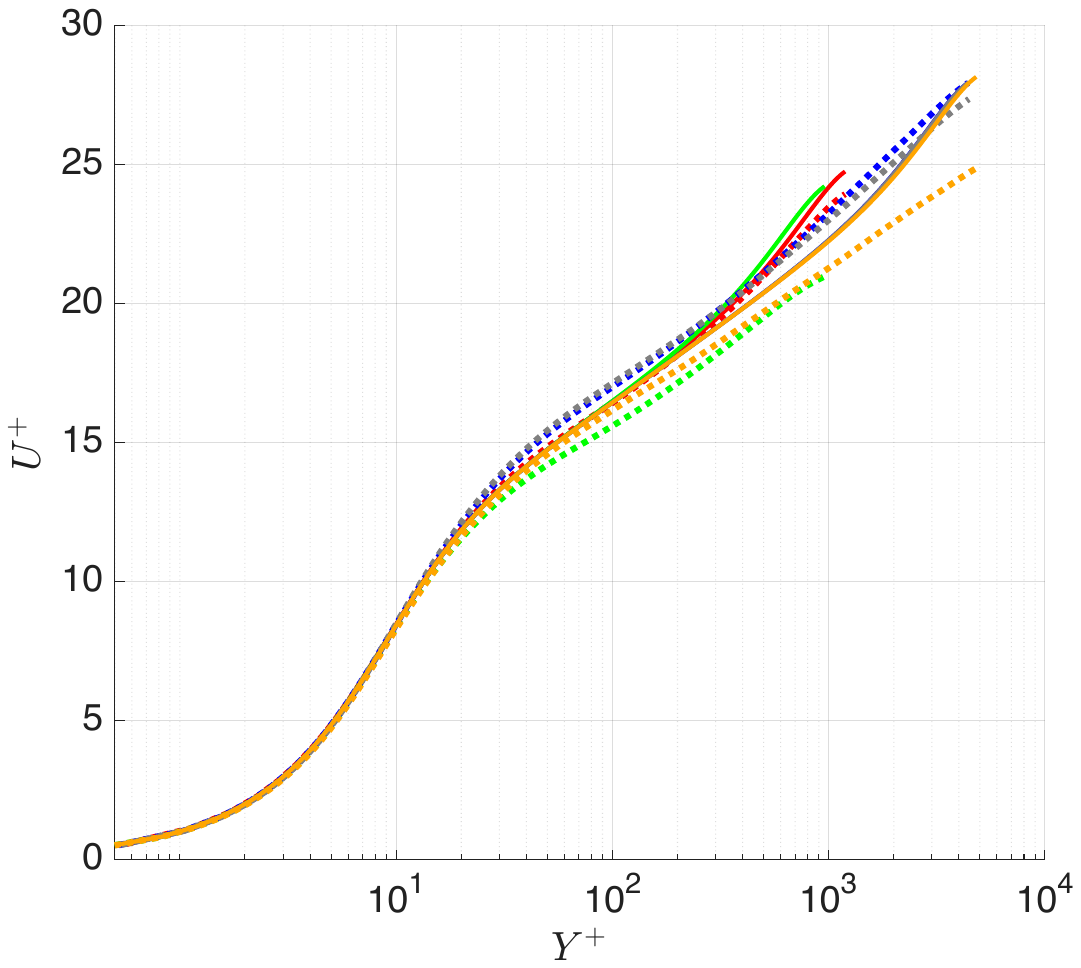}
    \caption{GFM \cite{Griffin2021}}
    \label{fig:consistency_gfm}
  \end{subfigure}
  \hfill
  \begin{subfigure}[b]{0.49\textwidth}
    \centering
    \includegraphics[width=\textwidth,trim={1.5cm 5cm 1.5cm 5cm},clip]{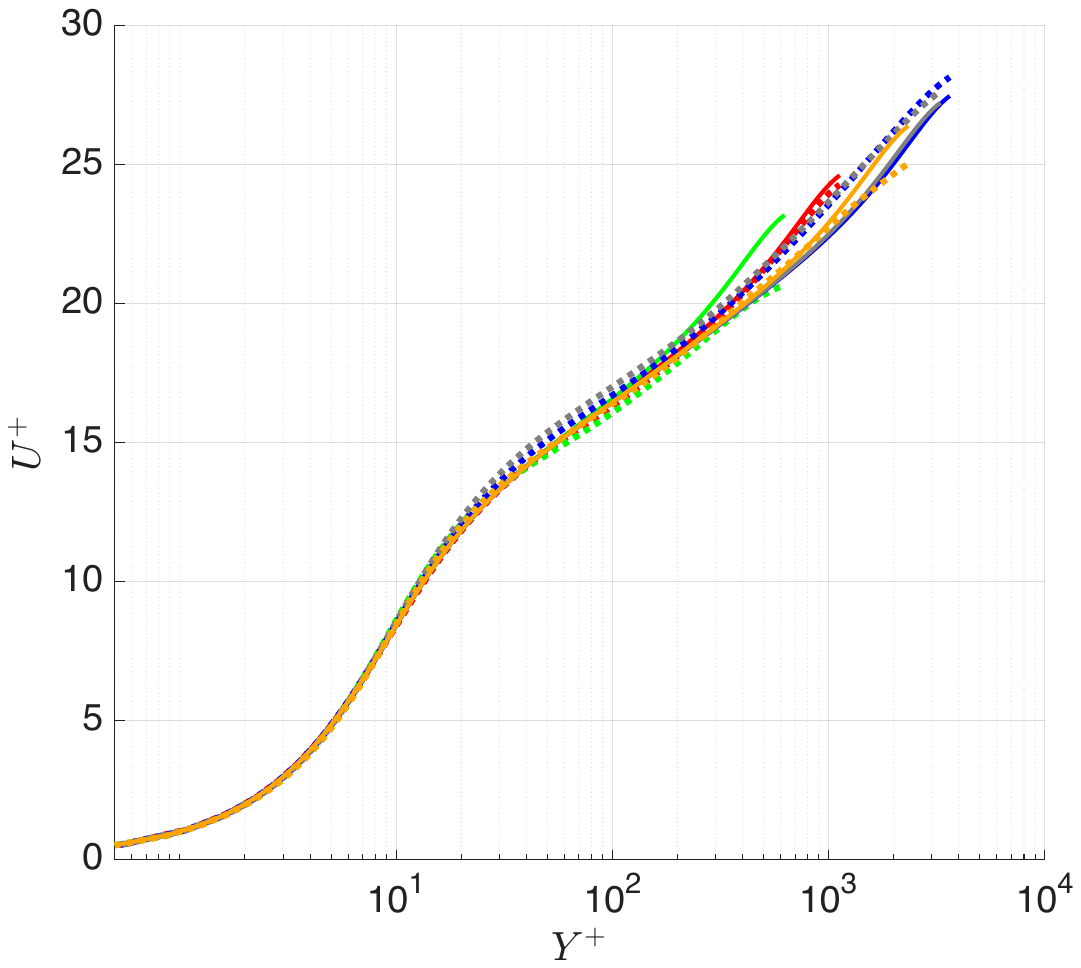}
    \caption{Volpiani \cite{volpiani2020}}
    \label{fig:consistency_volpiani}
  \end{subfigure}
  \hfill
  \begin{subfigure}[b]{0.49\textwidth}
    \centering
    \includegraphics[width=\textwidth,trim={1.5cm 5cm 1.5cm 5cm},clip]{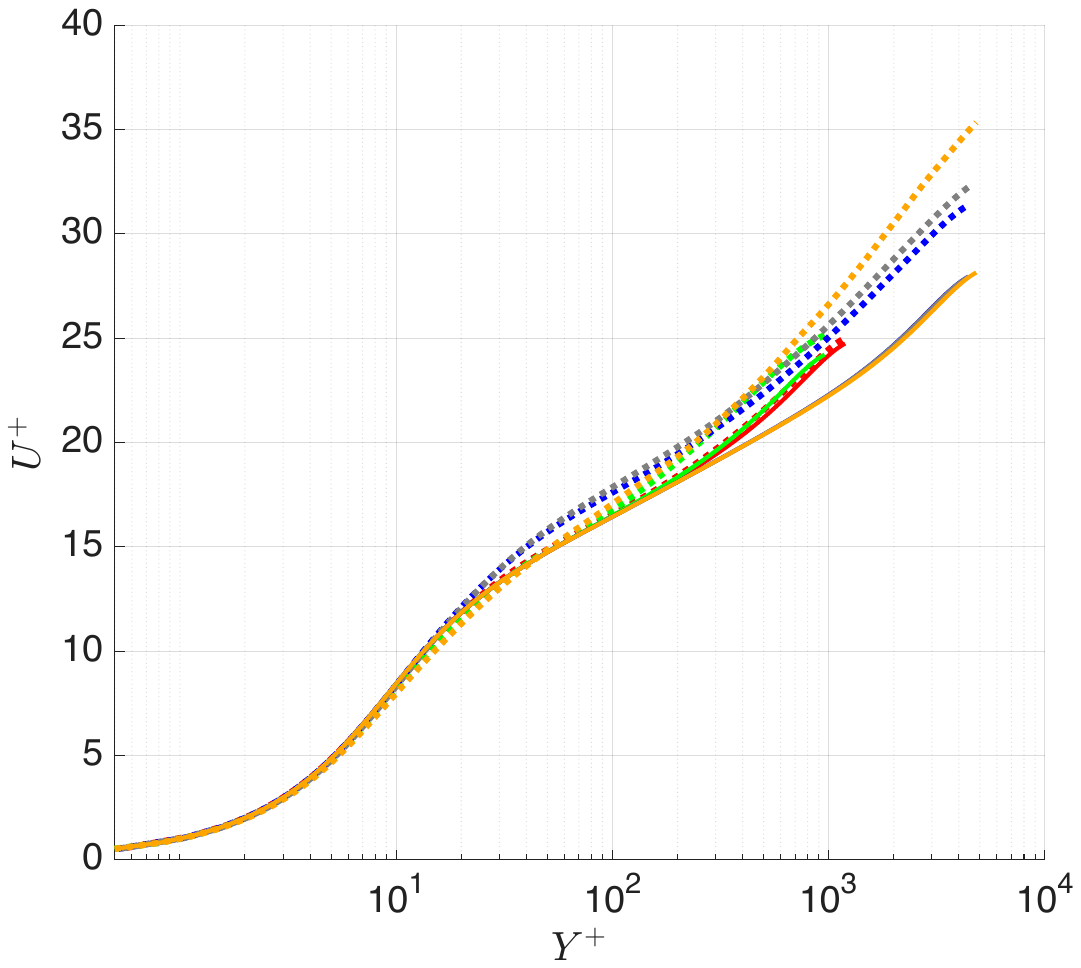}
    \caption{HLLP \cite{hasan2023}}
    \label{fig:consistency_hasan2023}
  \end{subfigure}

  \caption{Consistency assessment of existing compressibility transformations with respect to the incompressible inner--outer model in \Cref{sec:consistency}. Solid lines show the incompressible inner--outer model prediction $U_{\text{model}}^+(Y^+)$, and dotted lines show the transformed profiles $U_{\text{transformed}}^+(Y^+)$. Colors denote different hypersonic ZPG TBL cases in \cite{Zhang2018}: \legendline{red}~M2p5, \legendline{green}~M6Tw025, \legendline{blue}~M6Tw076, \legendline{gray}~M8Tw048, and \legendline{orange}~M14Tw018.}
  \label{fig:consistency_all}
\end{figure}

The corresponding profiles are shown in \Cref{fig:consistency_gfm,fig:consistency_volpiani,fig:consistency_hasan2023}. Solid lines show $U_{\text{model}}^+(Y^+)$, while dotted lines indicate the transformed profiles $U_{\text{transformed}}^+(Y^+)$. In each case, although the transformations are able to produce a qualitative collapse of the mean velocity profiles, the differences with respect to the incompressible model remain appreciable, particularly in the outer layer and for the cold-wall hypersonic cases. It is important to note that the GFM and Volpiani transformations do not explicitly model wake effects, unlike the HLLP transformation. Despite the lack of such a treatment, the Volpiani model performs the best among the three, while still exhibiting relatively large error values.

In the context of the present consistency requirement, recall that the incompressible model assumes $dU^+/dY^+=F(Y^+)$. Therefore, the modeled eddy viscosity and TKE production profiles are given by
\begin{equation}
    \mu_T^*=\frac{\mu_T}{\mu}=\frac{1}{F(Y^+)}-1,
    \qquad
    P^*=\frac{\mu P}{\tau_w^2}=F(Y^+)-F(Y^+)^2,
\end{equation}
which are Mach-independent. Hence, a transformation that satisfies the consistency condition on $Y^+$ will collapse not only the velocity profiles but also the eddy viscosity and TKE production profiles. Since the above-mentioned $U_{\text{transformed}}^+(Y^+)$ profiles have considerable errors with respect to $U_{\text{model}}^+(Y^+)$, these transformations fail to collapse the eddy viscosity and TKE production profiles as well, which was previously confirmed in \cite{danis2024transformation}. This indicates that none of these transformations achieves the level of agreement needed for a strictly Mach-independent incompressible representation of the entire compressible state, which motivates the development of the new forward transformation in \Cref{sec:forward}. 

\subsection{Performance of the proposed forward transformation}
\label{sec:results_forward}

The proposed forward transformation in \Cref{sec:forward} is first assessed in a casewise fashion by fitting the coefficients $A_1$, $A_2$, and $C_w$ independently for each compressible DNS case of Zhang et al.\ \cite{Zhang2018}. To maintain consistency between the forward and inverse transformations, the temperature $T$ is computed from the TV relation using the DNS mean velocity profile $u$. The density $\rho$ is then obtained by assuming an ideal gas in a ZPG boundary layer, and the viscosity $\mu$ is evaluated from Sutherland's law. The optimization problem in \Cref{eq:forward_opt_problem} is solved subject to the monotonicity constraint \Cref{eq:Yinc_monotone}, and the resulting coefficients are summarized in \Cref{tab:casewise_coeffs}.

\begin{table}[htbp]
  \centering
  \caption{Casewise optimized coefficients $A_1$, $A_2$, and $C_w$ in \Cref{eq:Yinc_def} for the five compressible DNS cases of Zhang et al.\ \cite{Zhang2018}.}
  \label{tab:casewise_coeffs}
  \begin{tabular}{lccc}
    \hlineB{3}
    Case      & $A_1$  & $A_2$  & $C_w$   \\
    \hline
    M2p5      & 0.3920 & 1.0000 & 0.2687 \\
    M6Tw025   & 0.2007 & 1.0000 & 0.9331 \\
    M6Tw076   & 0.2388 & 1.0000 & 0.6754 \\
    M8Tw048   & 0.2290 & 1.0000 & 0.9388 \\
    M14Tw018  & 0.1590 & 0.7934 & 1.0000 \\
    \hlineB{3}
  \end{tabular}
\end{table}

Across the five high-Mach number cases, $A_1$ decreases as wall cooling becomes stronger, indicating a growing contribution from the integral-type basis functions $Y_2^+$ and $Y_3^+$ in the cold-wall regime. The coefficient $A_2$ remains close to unity for most cases but deviates for the coldest case M14Tw018. Due to the lack of available data at higher Mach numbers with stronger wall cooling, how $A_2$ behaves in that regime is currently unclear. The wake coefficient $C_w$ increases toward unity as outer layer compressibility effects intensify, reaching $C_w \approx 1$ for the most strongly cooled case M14Tw018.

To enable predictions for new flow conditions without re-optimizing $\{A_1,A_2,C_w\}$, the casewise optimal coefficients in \Cref{tab:casewise_coeffs} are correlated with the wall-based parameters $(M_\tau,B_q)$ using the regression forms in \Cref{eq:A1_reg,eq:A2_reg,eq:Cw_reg}. Two models are considered: a multi-linear model based on the feature vector $\boldsymbol{\phi}_1$ in \Cref{eq:phi1_features}, and a multi-quadratic model based on $\boldsymbol{\phi}_2$ in \Cref{eq:phi2_features}. For each model, the regression coefficients are obtained by least squares fitting over the set of five DNS cases.

The accuracy of these regressions is evaluated by comparing the transformed velocity profiles obtained with the regressed coefficients to those obtained with the incompressible inner--outer model. The error measure is the maximum relative deviation from the incompressible inner--outer model,
\begin{equation}
    \varepsilon
    =\max_{Y^+\in[0,\delta_{i}^+]}\left|
    \frac{U_{\text{model}}^+(Y^+)-U_{\text{transformed}}^+(Y^+)}
         {U_{\text{model}}^+(Y^+)}
    \right|\times100,
\end{equation}
consistent with the definition used in \Cref{sec:consistency,sec:results_consistency}.

\Cref{tab:forward_errors} compares the maximum errors obtained using the casewise optimized model (without $(M_\tau,B_q)$ dependence) and those obtained using the multi-linear and multi-quadratic models (with explicit $(M_\tau,B_q)$ dependence). The multi-linear model reproduces the casewise performance to within about $0.3$ percentage points for all cases, with overall errors below $4\%$. The multi-quadratic model further improves the representation of the dependence on $(M_\tau,B_q)$. The regressed coefficients reproduce the casewise optimized performance to within numerical precision for all five cases; the casewise and regressed errors are indistinguishable at the reported accuracy. This outcome is expected, since the number of unknowns in the multi-quadratic fit (five) matches the number of training cases (five), so the regression effectively reduces to interpolation.

\begin{table}[htbp]
  \centering
  \caption{Maximum relative error $\varepsilon$ for the proposed forward transformations.}
  \label{tab:forward_errors}
  \begin{tabular}{lccc}
    \hlineB{3}
    Case      & $\varepsilon_{\text{casewise}}$ [\%] & $\varepsilon_{\text{multi-linear}}$ [\%] & $\varepsilon_{\text{multi-quadratic}}$ [\%] \\
    \hline
    M2p5      & 0.992 & 1.491 & 0.992 \\
    M6Tw025   & 2.221 & 2.474 & 2.221 \\
    M6Tw076   & 1.623 & 1.853 & 1.623 \\
    M8Tw048   & 1.993 & 2.155 & 1.993 \\
    M14Tw018  & 3.679 & 3.743 & 3.679 \\
    \hlineB{3}
  \end{tabular}
\end{table}

For the multi-linear model, the explicit dependence of the transformed coordinate $Y_{\text{inc}}^+$ on $(M_\tau,B_q)$ can be written in closed form. With
\begin{equation}
  \boldsymbol{\phi}_1 =
  \begin{bmatrix}
    M_{\tau} \\
    B_q
  \end{bmatrix},
\end{equation}
the regression relations in \Cref{eq:A1_reg,eq:A2_reg,eq:Cw_reg} become
\begin{equation}
  A_1(M_{\tau},B_q) = \exp\!\left(\boldsymbol{\phi}_1^\top \boldsymbol{\theta}_{A_1}\right),
  \qquad
  A_2(M_{\tau},B_q) = \exp\!\left(\boldsymbol{\phi}_1^\top \boldsymbol{\theta}_{A_2}\right),
  \qquad
  C_w(M_{\tau},B_q) = \boldsymbol{\phi}_1^\top \boldsymbol{\theta}_{C_w},
\end{equation}
where the fitted coefficient vectors are
\begin{equation}
  \boldsymbol{\theta}_{A_1}
  = \begin{bmatrix} -10.797 \\[2pt] 1.6859 \end{bmatrix},
  \qquad
  \boldsymbol{\theta}_{A_2}
  = \begin{bmatrix} 0.33923 \\[2pt] -1.2176 \end{bmatrix},
  \qquad
  \boldsymbol{\theta}_{C_w}
  = \begin{bmatrix} 5.0506 \\[2pt] 0.42472 \end{bmatrix}.
\end{equation}

For the multi-quadratic model, the feature vector is extended to
\begin{equation}
  \boldsymbol{\phi}_2 =
  \begin{bmatrix}
    M_{\tau} \\
    B_q      \\
    M_{\tau}^2 \\
    M_{\tau} B_q \\
    B_q^2
  \end{bmatrix},
\end{equation}
and the same functional forms \Cref{eq:A1_reg,eq:A2_reg,eq:Cw_reg} are used, with
\begin{equation}
  A_1(M_{\tau},B_q) = \exp\!\left(\boldsymbol{\phi}_2^\top \boldsymbol{\theta}_{A_1}\right),
  \qquad
  A_2(M_{\tau},B_q) = \exp\!\left(\boldsymbol{\phi}_2^\top \boldsymbol{\theta}_{A_2}\right),
  \qquad
  C_w(M_{\tau},B_q) = \boldsymbol{\phi}_2^\top \boldsymbol{\theta}_{C_w},
\end{equation}
where the fitted coefficient vectors are
\begin{equation}
  \boldsymbol{\theta}_{A_1}
  = \begin{bmatrix}
      -8.5109 \\[2pt]
      -12.478 \\[2pt]
      -30.180 \\[2pt]
      178.77  \\[2pt]
      -95.161
    \end{bmatrix},
  \qquad
  \boldsymbol{\theta}_{A_2}
  = \begin{bmatrix}
       0.95781 \\[2pt]
      11.683   \\[2pt]
     -11.662   \\[2pt]
     -61.892   \\[2pt]
       2.301
    \end{bmatrix},
  \qquad
  \boldsymbol{\theta}_{C_w}
  = \begin{bmatrix}
      -0.90151 \\[2pt]
       3.9064  \\[2pt]
      48.005   \\[2pt]
     -30.572   \\[2pt]
      -8.6027
    \end{bmatrix}.
\end{equation}

Together with \Cref{eq:Y1_def,eq:Y2_def,eq:Y3_def}, inserting these results into \Cref{eq:Yinc_def} provides a compact, parameterized closed-form expression for the forward transformation that achieves $1\%$--$4\%$ consistency with the incompressible inner--outer model across the hypersonic database considered:
\begin{equation}
    \begin{aligned}
        Y_{\text{inc}}^+(y;M_\tau,B_q) &= A_1(M_{\tau},B_q)\,Y_1^+(y) 
        + \bigl[1-A_1(M_{\tau},B_q)\bigr]\Bigl\{Y_2^+(y) - A_2(M_{\tau},B_q)\,Y_3^+(y)\Bigr\},\\
        Y_1^+(y) &= \frac{\rho(y)\,u_\tau^*(y)}{\mu(y)}\,y,\\
        Y_2^+(y) &= \int_0^{y} \frac{\rho(y')\,u_\tau^*(y')}{\mu(y')}\,dy',\\
        Y_3^+(y) &= \int_0^{y} \frac{\rho(y')\,u_q}{\mu(y')}\,dy'.
    \end{aligned}
  \label{eq:Yinc_closed}
\end{equation}
For illustration, the multi-linear model can be written approximately as
\begin{equation}
    \begin{aligned}
        Y_{\text{inc}}^+(y;M_\tau,B_q)
        &\approx \exp{\!\bigl(-10.8\,M_{\tau}+1.69\,B_q\bigr)}\,Y_1^+(y) \\
        &\quad+ \Bigl[1-\exp{\!\bigl(-10.8\,M_{\tau}+1.69\,B_q\bigr)}\Bigr]
        \Bigl\{Y_2^+(y) - \exp{\!\bigl(0.34\,M_{\tau}-1.22\,B_q\bigr)}\,Y_3^+(y)\Bigr\},
    \end{aligned}
\end{equation}
with the corresponding mean velocity transformation
\begin{equation}
  U_{\text{transformed}}^+(Y_{\text{inc}}^+)
  \approx \int_0^{Y_{\text{inc}}^+} 
    \left[
      \frac{\mu}{\tau_w}\,\frac{du}{dy}
      + \left(5.05\,M_\tau + 0.42\,B_q\right)\,F_{\text{outer}}(Y';\delta_i^+)
    \right]\,dY'.
\end{equation}

\begin{figure}[htbp]
  \centering

  \begin{subfigure}[b]{0.49\textwidth}
    \centering
    \includegraphics[width=\textwidth,trim={1.5cm 5cm 1.5cm 5cm},clip]{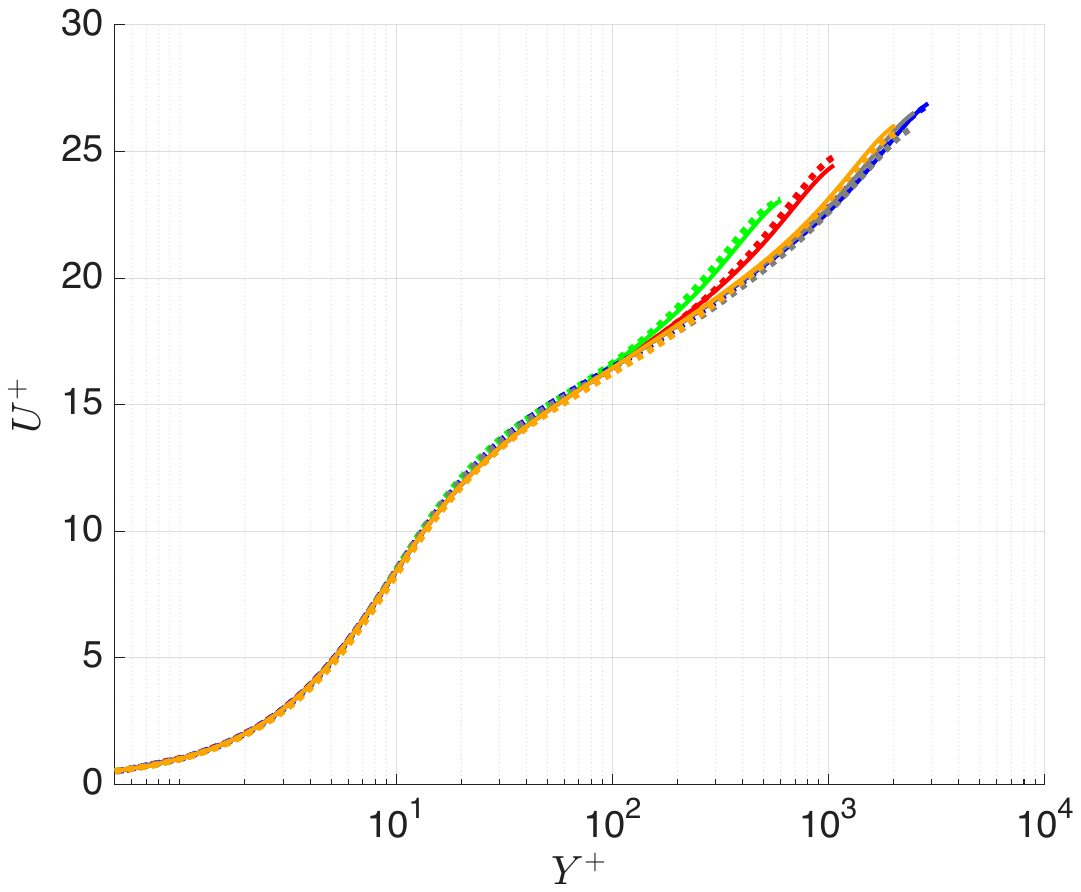}
    \caption{Multi-linear fit}
    \label{fig:forward_linear}
  \end{subfigure}
  \hfill
  \begin{subfigure}[b]{0.49\textwidth}
    \centering
    \includegraphics[width=\textwidth,trim={1.5cm 5cm 1.5cm 5cm},clip]{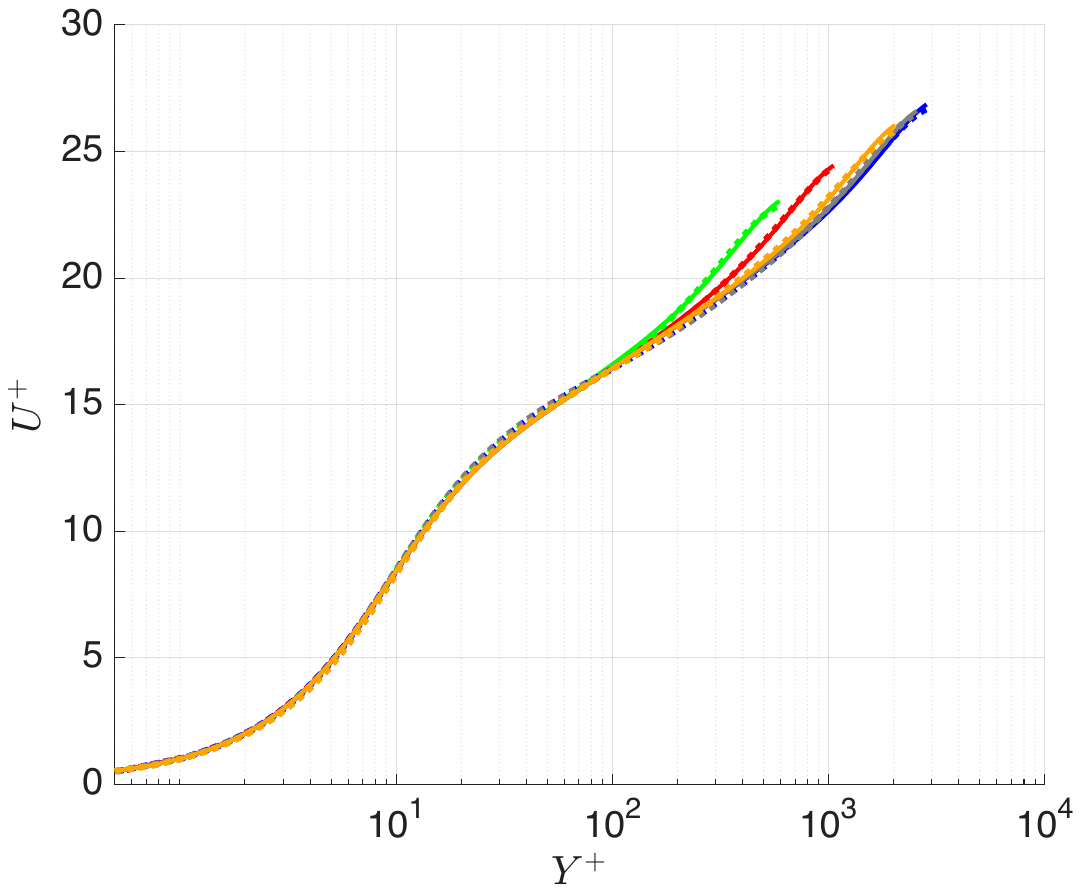}
    \caption{Multi-quadratic fit}
    \label{fig:forward_quadratic}
  \end{subfigure}

  \caption{Transformed velocity profiles obtained with the proposed forward transformation using (a) the multi-linear and (b) the multi-quadratic fit in $(M_\tau,B_q)$. Solid lines show the incompressible inner--outer model $U_{\text{model}}^+(Y^+_{\text{inc}})$, and dotted lines show the transformed profiles $U_{\text{transformed}}^+(Y^+_\text{inc})$ with the corresponding regression-based coefficients. Colors denote different hypersonic ZPG TBL cases in \cite{Zhang2018}: \legendline{red}~M2p5, \legendline{green}~M6Tw025, \legendline{blue}~M6Tw076, \legendline{gray}~M8Tw048, and \legendline{orange}~M14Tw018.}
  \label{fig:forward_regression}
\end{figure}

The practical impact of the multi-linear and multi-quadratic fits is illustrated in \Cref{fig:forward_regression}, which compares the transformed velocity profiles obtained with the corresponding regression-based coefficients to those obtained with the incompressible inner--outer model using the same $Y_{\text{inc}}^+$. The results show that the proposed forward transformation collapses $U_{\text{transformed}}^+(Y_{\text{inc}}^+)$ onto $U_{\text{model}}^+(Y_{\text{inc}}^+)$ with excellent accuracy, thereby satisfying the consistency requirement.

\subsection{Performance of the inverse transformation}
\label{sec:results_inverse}

First, the mean shear model is examined for the inverse transformation. Recall that $du/dy$ satisfies
\begin{equation}
  \frac{\mu}{\tau_w}\,\frac{du}{dy}
  = F_{\text{inner}}(Y_{\text{inc}}^+)
    + \bigl(1- C_w\bigr)\,F_{\text{outer}}\!\left(Y_{\text{inc}}^+;\delta_i^+\right),
\end{equation}
and that $C_w \to 1$ as the Mach number and wall cooling increase (see \Cref{tab:casewise_coeffs}). This implies that $1-C_w \to 0$, and thus
\begin{equation}
  \frac{\mu(y)}{\tau_w}\,\frac{du}{dy}
  \;\longrightarrow\;
  F_{\text{inner}}(Y_{\text{inc}}^+).
\end{equation}
In other words, in the high Mach-number, cold-wall limit, the transformed shear is controlled entirely by the inner layer model. This observation is consistent with the trends reported in \cite{Ying_Li_Fu_2025}, where the mean velocity was also found to approach the inner layer profile, $u \to u_{\text{inner}}$, under the same limiting conditions. On the other hand, in the incompressible limit $C_w\to 0$ (or equivalently $1-C_w \to 1$), and thus
\begin{equation}
    \frac{\mu}{\tau_w}\,\frac{du}{dy}
    \;\longrightarrow\;
    F_{\text{inner}}(Y_{\text{inc}}^+)
    + F_{\text{outer}}\!\left(Y_{\text{inc}}^+;\delta_i^+\right).
\end{equation}
That is, the wake is modeled solely by the incompressible outer model, as expected.

The proposed inverse (incompressible-to-compressible) transformation of \Cref{sec:inverse} is next assessed at the DNS sampling station of Zhang et al.\ \cite{Zhang2018}. For each hypersonic case, the inverse solver is driven by the prescribed freestream and wall conditions, the calibrated forward model, and the boundary layer thickness $\delta$ corresponding to the DNS measurement station.

We consider three variants: the present inverse model with the multi-linear fit, the multi-quadratic fit, and the inverse HLLP transformation \cite{hasan2024} as a reference. Note that the HLLP transformation is also implemented within the present inverse solver framework of \Cref{fig:workflow_inverse}. This corresponds to setting $Y_{\text{inc}}^+=y^*$ and evaluating $du/dy$ using Eq.~(3) of Hasan et al.\ \cite{hasan2024}, with the appropriate $Re_\theta$-based adjustment for the wake parameter $\Pi$.

\Cref{tab:inverse_all} summarizes the performance of the proposed inverse transformations (along with HLLP) in terms of several boundary layer parameters, including the maximum relative errors in velocity ($u$--err [\%]) and temperature ($T$--err [\%]) over the entire boundary layer:
\begin{equation}
    \begin{aligned}
        u\text{--err [\%]} &= \max_{y\in[0,\delta]}\left|\frac{u_{\text{transformed}}(y)-u_{\text{DNS}}(y)}{u_{\text{DNS}}(y)} \right|\times100,\\
        T\text{--err [\%]} &= \max_{y\in[0,\delta]}\left|\frac{T_{\text{transformed}}(y)-T_{\text{model}}(y)}{T_{\text{model}}(y)} \right|\times100.
    \end{aligned}
\end{equation}
Since temperature profiles in both forward and inverse models are obtained from the TV relation, the error in reconstructed temperature is computed with respect to $T_{\text{model}}(y)$, which is obtained from the same TV relation in \Cref{eq:zhang-temperature} using the DNS velocity $u_{\text{DNS}}(y)$.

Both versions of the present inverse model recover the key boundary layer parameters with good accuracy. Recall that the multi-linear model is obtained via regression, whereas the multi-quadratic model effectively corresponds to an interpolation over the five training cases. Despite this reduction in approximation order, the multi-linear model yields reasonable accuracy in the reconstructed boundary layer parameters. For most quantities, the predicted values match the corresponding DNS results within only a few percent. The maximum relative error in velocity remains below $3.4\%$ for the multi-linear model and $2.9\%$ for the multi-quadratic model. Both models attain higher error levels in temperature, but still provide better accuracy than the HLLP model in most cases.

\begin{table}[htbp]
  \centering
  \caption{Inverse reconstruction at the DNS measurement station.
  DNS values from \cite{Zhang2018} are compared with the present inverse models
  (multi-linear and multi-quadratic fits in $(M_\tau,B_q)$) and the inverse HLLP
  transformation \cite{hasan2024}. For each case, the first row lists DNS values,
  and subsequent rows list model predictions.}
  \label{tab:inverse_all}
  \begin{tabular}{llccccccrr}
    \hlineB{3}
    Case & Model
      & $Re_\tau$
      & $Re_\theta$
      & $y_\tau\,[\mu\mathrm{m}]$
      & $u_\tau\,[\mathrm{m/s}]$
      & $B_q$
      & $M_\tau$
      & $u$--err [\%]
      & $T$--err [\%] \\
    \hline
    M2p5
      & DNS              & 510 & 2835  & 15.0  & 40.60 & 0.00 & 0.08 &   -- &   -- \\
      & Multi-linear     & 511 & 2791  & 14.9  & 40.92 & 0.00 & 0.09 & 2.36 & 1.48 \\
      & Multi-quadratic  & 504 & 2867  & 15.1  & 40.36 & 0.00 & 0.08 & 1.16 & 0.59 \\
      & HLLP \cite{hasan2024}   & 495 & 2888  & 15.4  & 39.67 & 0.00 & 0.08 & 3.78 & 1.64 \\
    \hline
    M6Tw025
      & DNS              & 450 & 2121  &  8.0  & 33.80 & 0.14 & 0.17 &   -- &   -- \\
      & Multi-linear     & 454 & 2180  &  8.0  & 34.07 & 0.13 & 0.17 & 2.70 & 4.29 \\
      & Multi-quadratic  & 450 & 2182  &  8.0  & 33.73 & 0.13 & 0.17 & 2.35 & 4.33 \\
      & HLLP \cite{hasan2024}   & 453 & 2122  &  8.0  & 33.94 & 0.13 & 0.17 & 3.42 & 1.03 \\
    \hline
    M6Tw076
      & DNS              & 453 & 9455  & 52.6  & 45.10 & 0.02 & 0.13 &   -- &   -- \\
      & Multi-linear     & 464 & 9926  & 51.5  & 44.30 & 0.02 & 0.13 & 1.76 & 3.89 \\
      & Multi-quadratic  & 462 & 9939  & 51.7  & 44.15 & 0.02 & 0.13 & 1.98 & 4.01 \\
      & HLLP \cite{hasan2024}   & 465 & 9932  & 51.3  & 44.46 & 0.02 & 0.13 & 4.29 & 7.82 \\
    \hline
    M8Tw048
      & DNS              & 480 & 9714  & 73.5  & 54.30 & 0.06 & 0.15 &   -- &   -- \\
      & Multi-linear     & 476 & 10685 & 74.1  & 53.07 & 0.06 & 0.15 & 3.54 & 5.73 \\
      & Multi-quadratic  & 481 & 10466 & 73.4  & 53.64 & 0.06 & 0.16 & 2.24 & 2.86 \\
      & HLLP \cite{hasan2024}   & 486 & 10484 & 72.7  & 54.14 & 0.06 & 0.16 & 3.75 & 7.85 \\
    \hline
    M14Tw018
      & DNS              & 646 & 14408 & 102.4 & 67.60 & 0.19 & 0.19 &   -- &   -- \\
      & Multi-linear     & 639 & 14118 & 103.3 & 66.61 & 0.18 & 0.19 & 2.25 & 6.31 \\
      & Multi-quadratic  & 642 & 14200 & 102.9 & 66.86 & 0.18 & 0.19 & 2.87 & 8.03 \\
      & HLLP \cite{hasan2024}   & 661 & 13522 &  99.8 & 68.90 & 0.18 & 0.20 &10.31 &11.08 \\
    \hlineB{3}
  \end{tabular}
\end{table}

\begin{figure}[htbp]
  \centering
  \begin{subfigure}[b]{0.48\textwidth}
    \centering
    \includegraphics[width=1.00\textwidth,trim={1.5cm 6cm 1.5cm 6.5cm},clip]{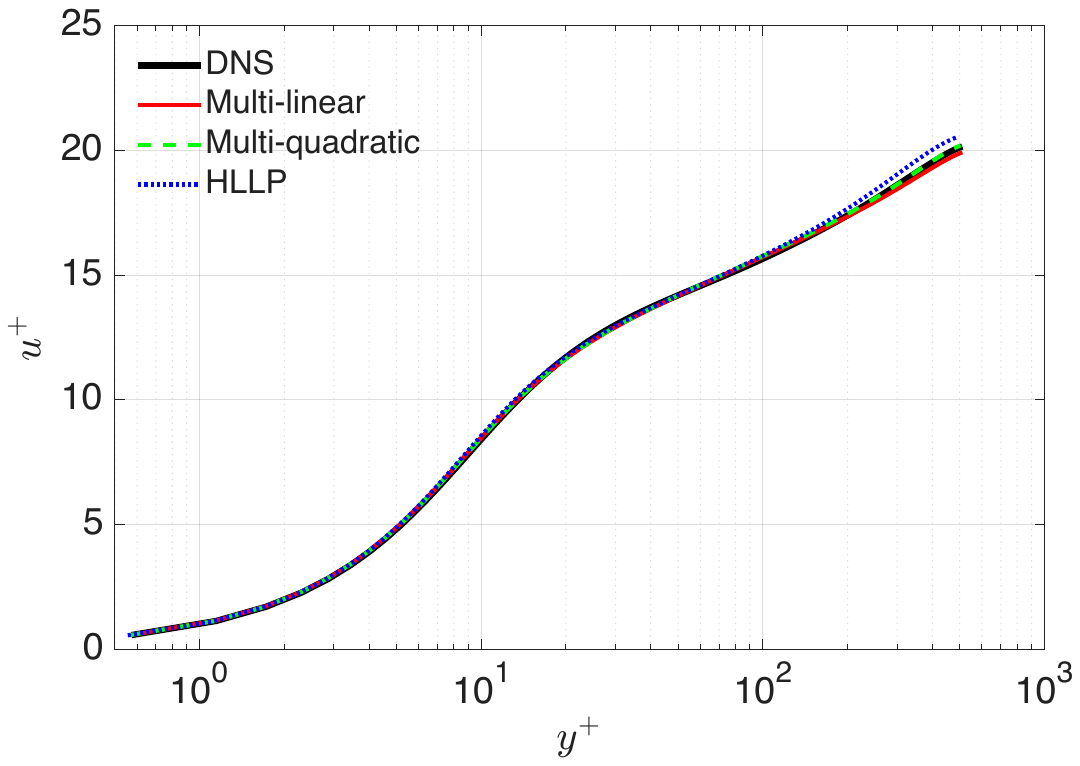}
    \caption{M2p5}
  \end{subfigure}
  \begin{subfigure}[b]{0.48\textwidth}
    \centering
    \includegraphics[width=1.00\textwidth,trim={1.5cm 6cm 1.5cm 6.5cm},clip]{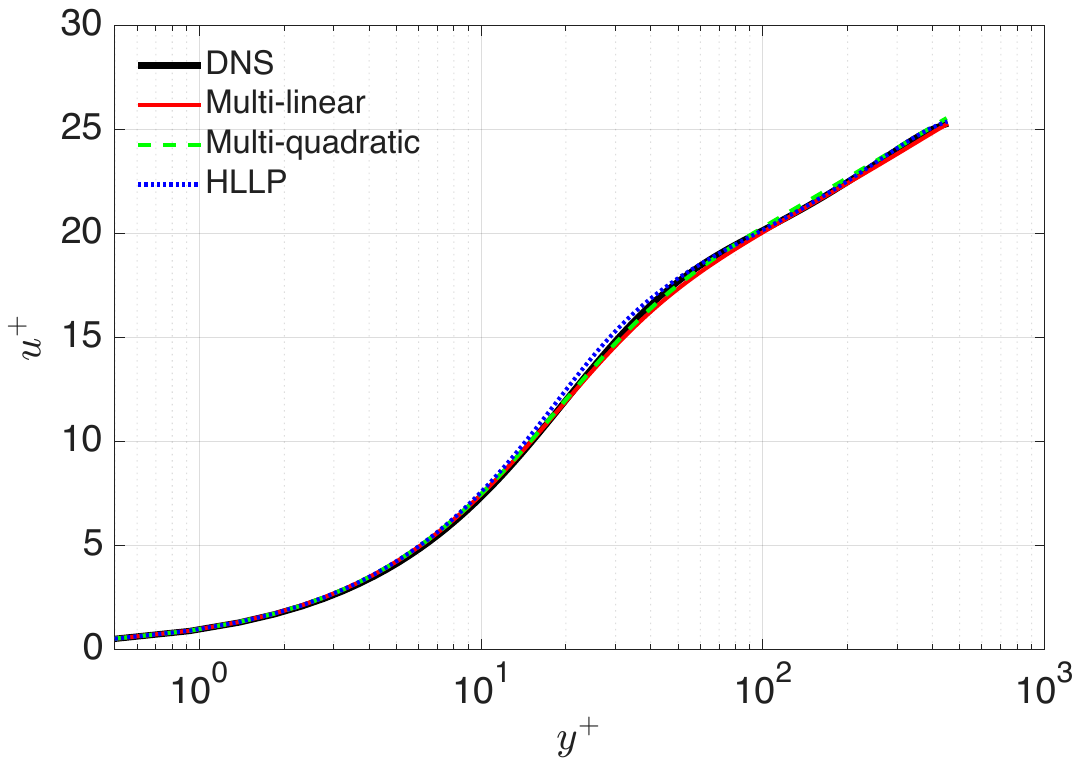}
    \caption{M6Tw025}
  \end{subfigure}
  \begin{subfigure}[b]{0.48\textwidth}
    \centering
    \includegraphics[width=1.00\textwidth,trim={1.5cm 6cm 1.5cm 6.5cm},clip]{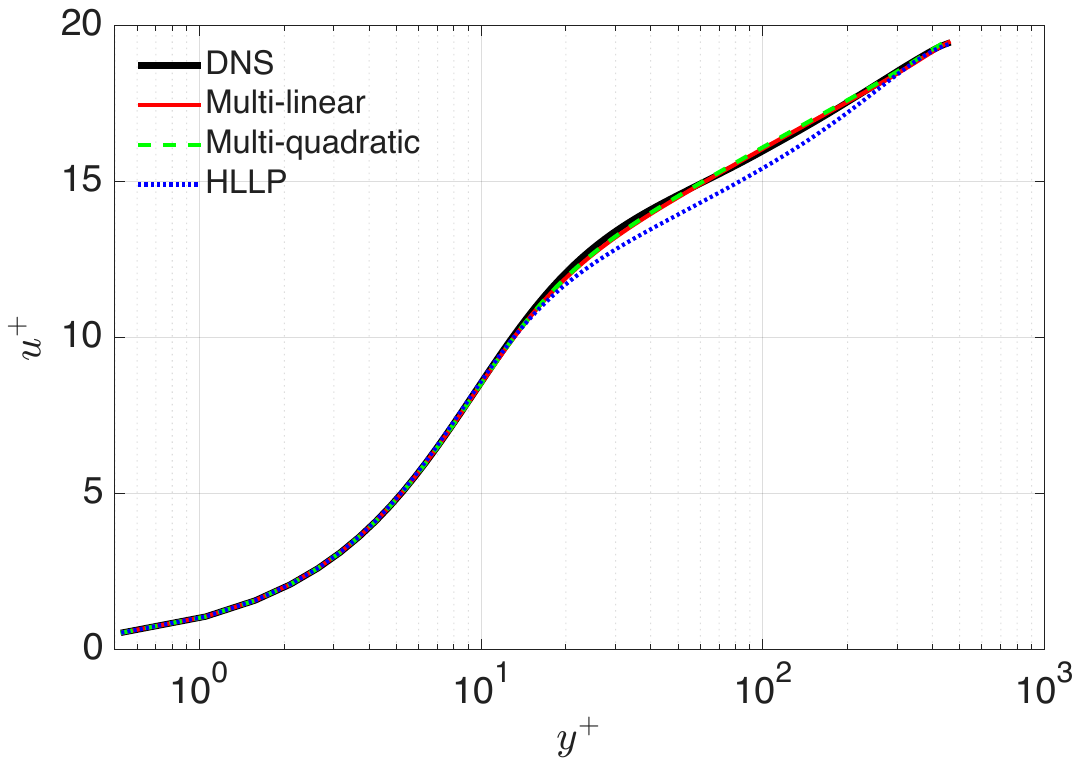}
    \caption{M6Tw076}
  \end{subfigure}
  \begin{subfigure}[b]{0.48\textwidth}
    \centering
    \includegraphics[width=1.00\textwidth,trim={1.5cm 6cm 1.5cm 6.5cm},clip]{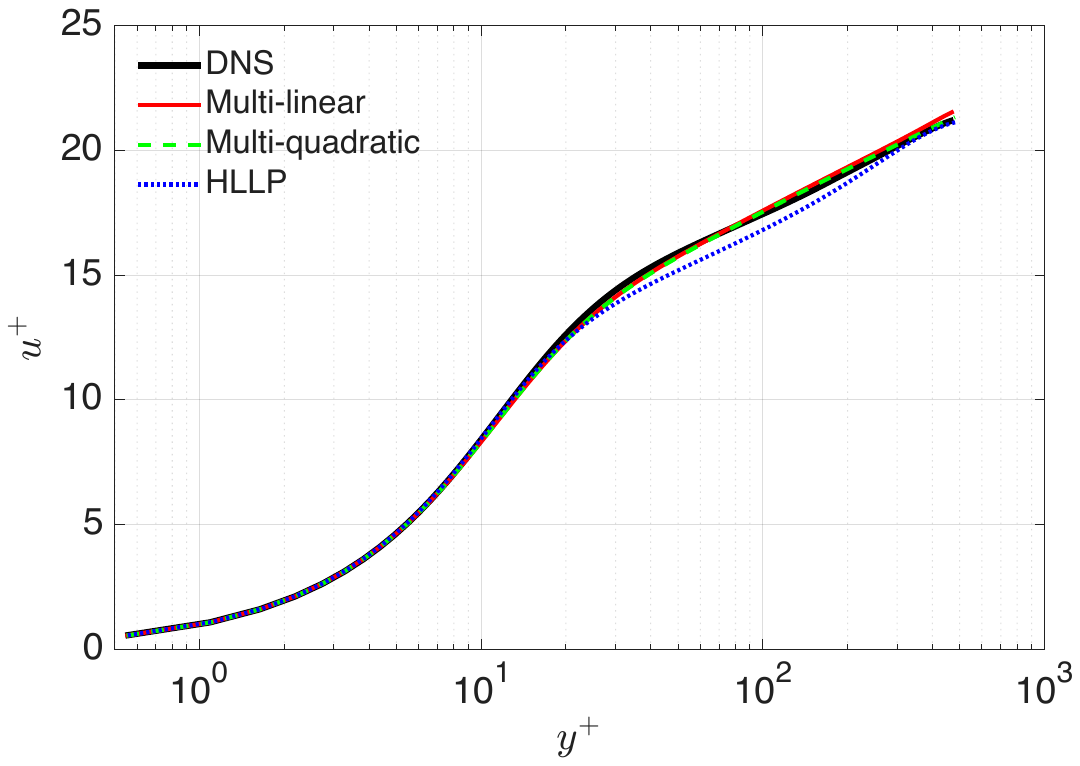}
    \caption{M8Tw048}
  \end{subfigure}
  \begin{subfigure}[b]{0.48\textwidth}
    \centering
    \includegraphics[width=1.00\textwidth,trim={1.5cm 6cm 1.5cm 6.5cm},clip]{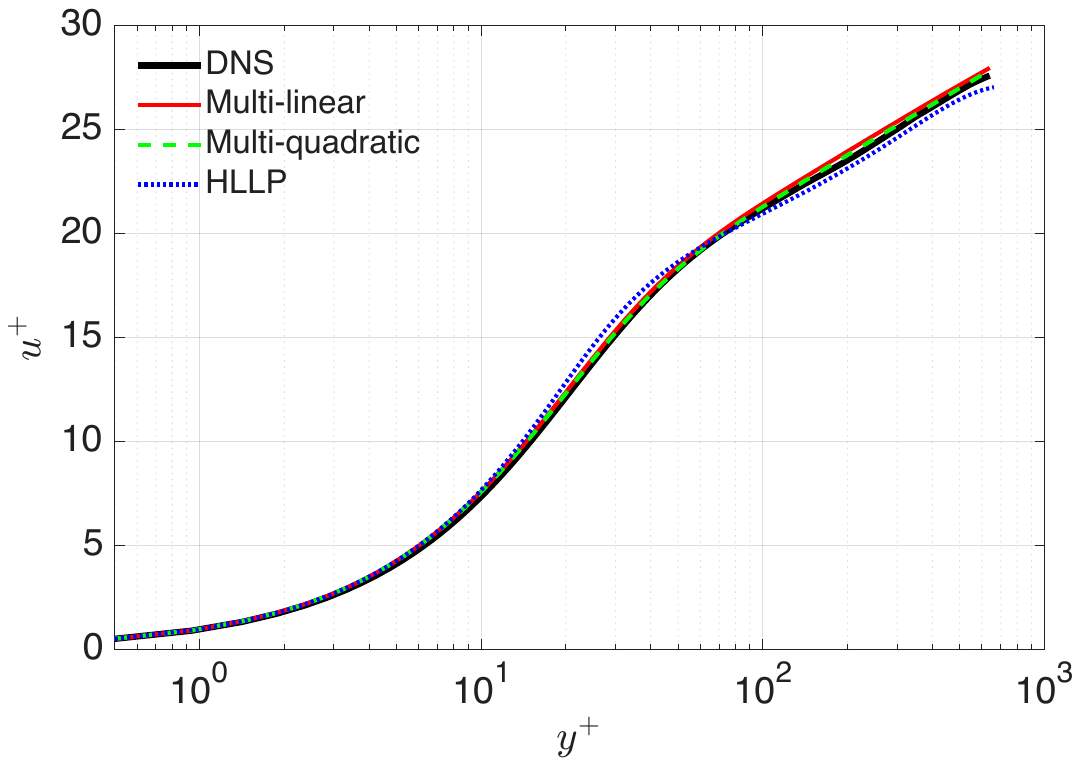}
    \caption{M14Tw018}
  \end{subfigure}
  \caption{Inverse reconstruction of mean velocity profiles at the DNS sampling
  station.}
  \label{fig:inverse_u_all}
\end{figure}

\begin{figure}[htbp]
  \centering
  \begin{subfigure}[b]{0.48\textwidth}
    \centering
    \includegraphics[width=1.00\textwidth,trim={1.25cm 6cm 1.5cm 6.5cm},clip]{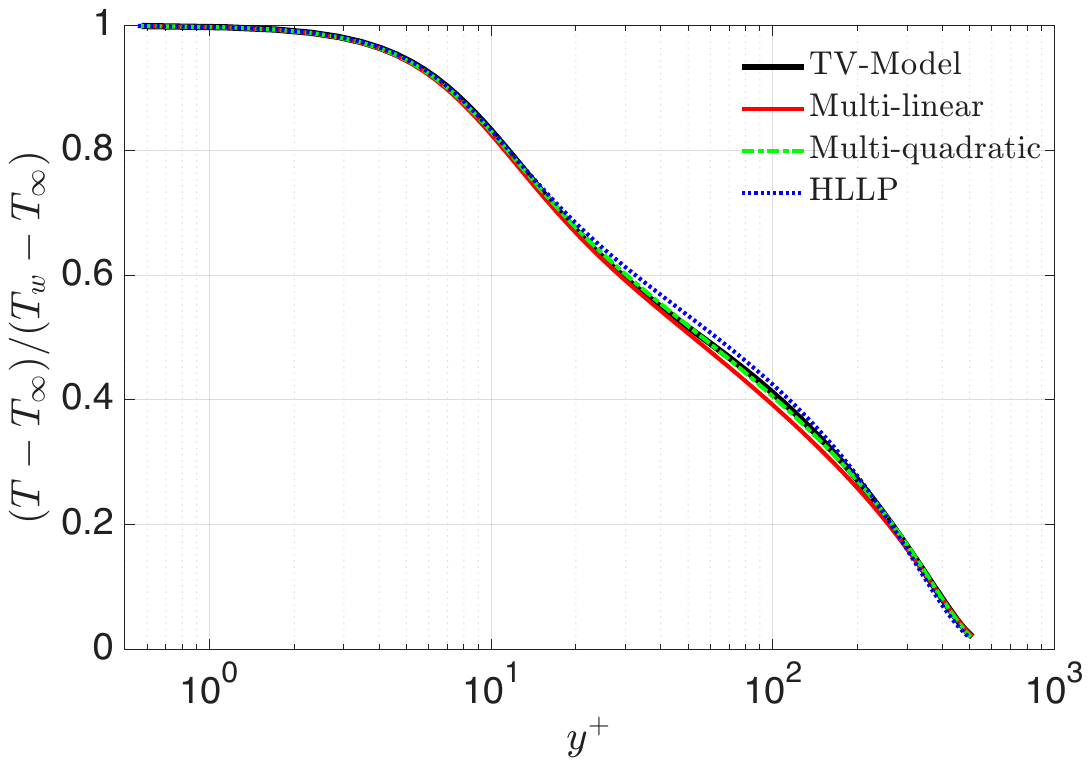}
    \caption{M2p5}
  \end{subfigure}
  \hfill
  \begin{subfigure}[b]{0.48\textwidth}
    \centering
    \includegraphics[width=1.00\textwidth,trim={1.25cm 6cm 1.5cm 6.5cm},clip]{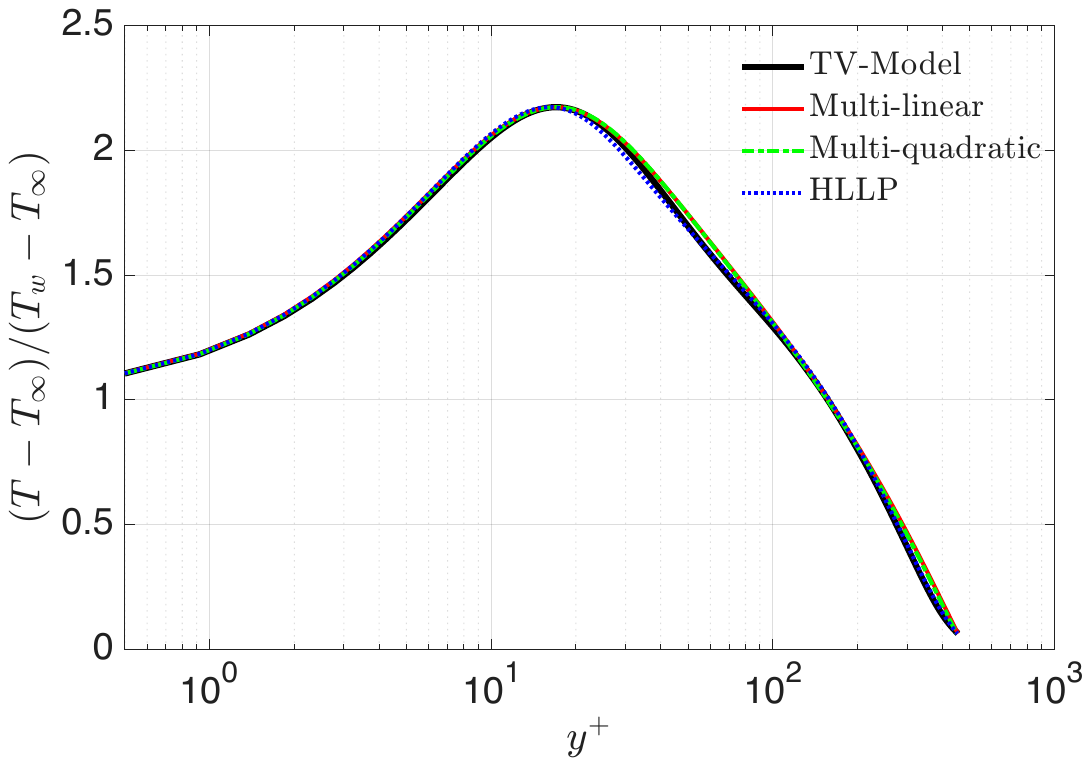}
    \caption{M6Tw025}
  \end{subfigure}
  \begin{subfigure}[b]{0.48\textwidth}
    \centering
    \includegraphics[width=1.00\textwidth,trim={1.25cm 6cm 1.5cm 6.5cm},clip]{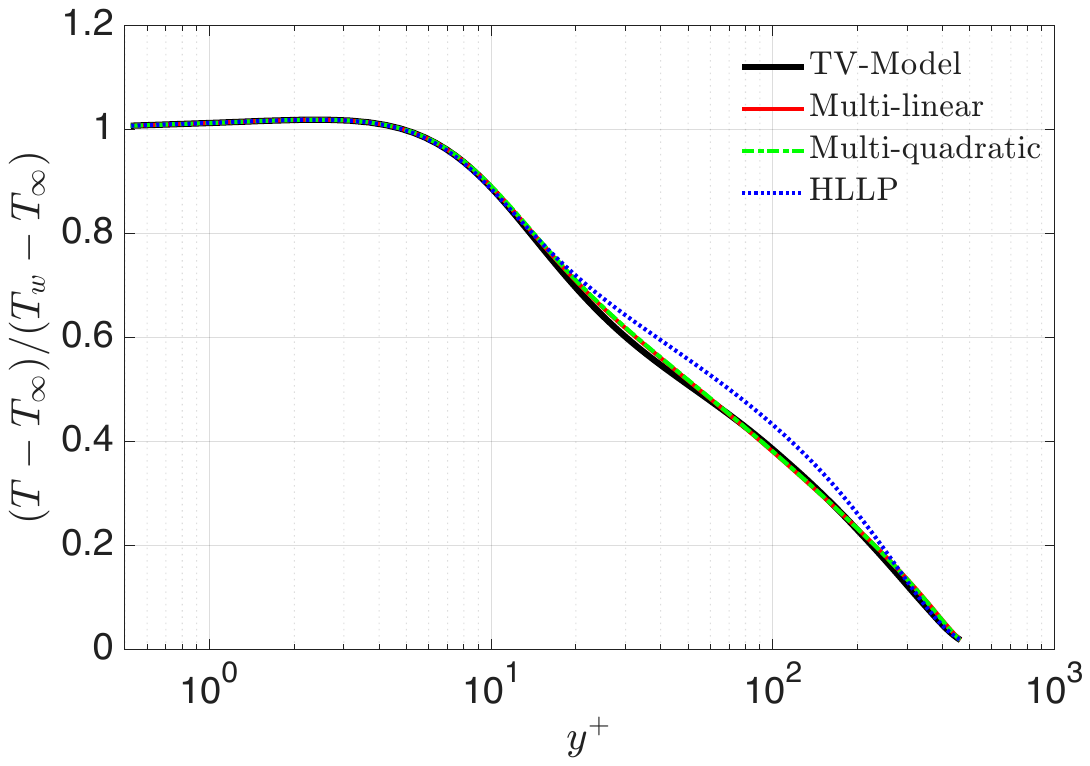}
    \caption{M6Tw076}
  \end{subfigure}
  \hfill
  \begin{subfigure}[b]{0.48\textwidth}
    \centering
    \includegraphics[width=1.00\textwidth,trim={1.25cm 6cm 1.5cm 6.5cm},clip]{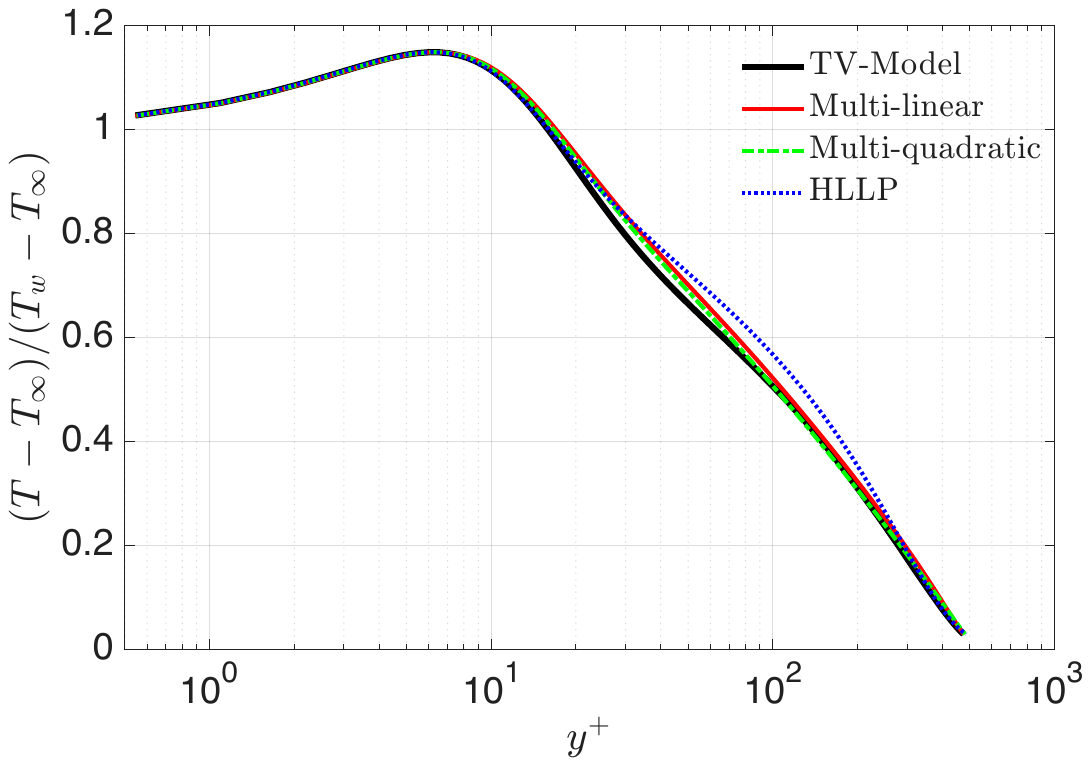}
    \caption{M8Tw048}
  \end{subfigure}
  \begin{subfigure}[b]{0.48\textwidth}
    \centering
    \includegraphics[width=1.00\textwidth,trim={1.25cm 6cm 1.5cm 6.5cm},clip]{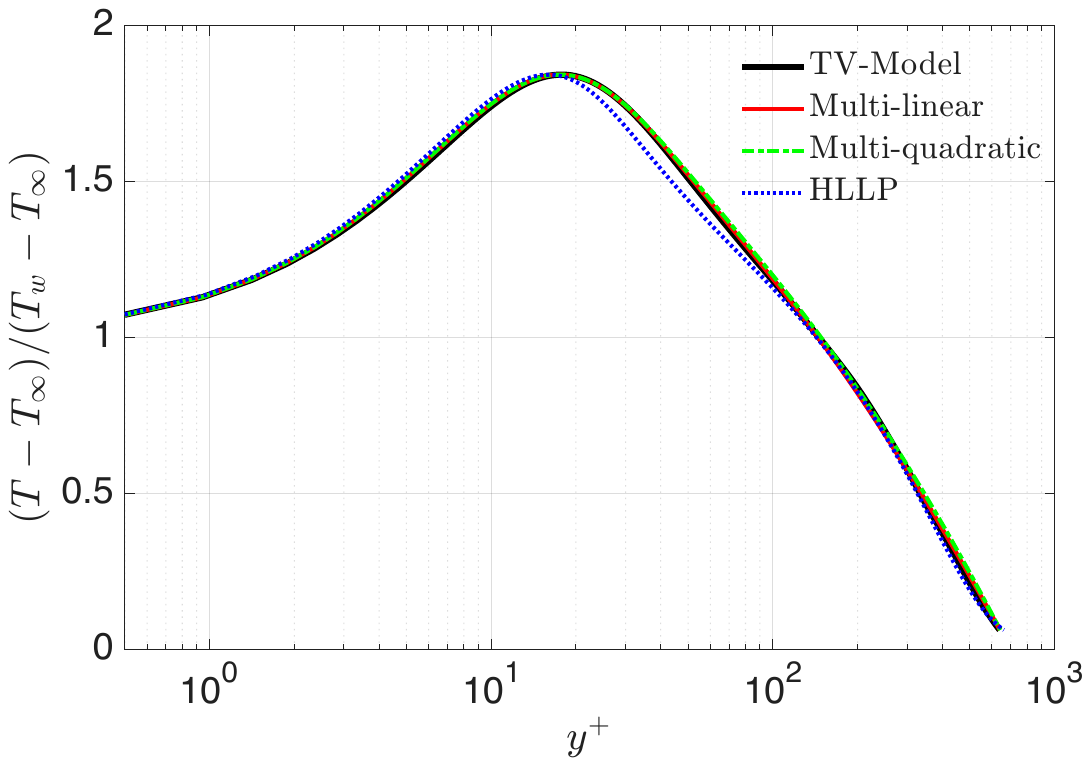}
    \caption{M14Tw018}
  \end{subfigure}
  \caption{Inverse reconstruction of mean temperature profiles at the DNS sampling
  station.}
  \label{fig:inverse_T_all}
\end{figure}

\Cref{fig:inverse_u_all,fig:inverse_T_all} show the predicted mean velocity and temperature profiles. In all cases, both proposed models remain in very good agreement with $u_{\text{DNS}}(y)$ and $T_{\text{model}}(y)$, whereas the HLLP model exhibits noticeable discrepancies, particularly for the M6Tw076 and M8Tw048 cases. It is also worth noting that the multi-linear (regressed) model is nearly as accurate as the multi-quadratic (interpolated) model for velocity and temperature predictions.

Until now, the proposed inverse models have been examined within the same parameter space (interpolation) in which the forward model was calibrated. One might argue that predictions outside this space (extrapolation) are even more important. However, the available hypersonic ZPG TBL data are limited, which constrains further tuning of the model coefficients. A pragmatic way to probe extrapolative behaviour is to keep the freestream and wall conditions fixed, while varying the boundary layer thickness $\delta$ and using the inverse solver to reconstruct the corresponding mean states. In this way, profiles of $C_f$ and $C_h$ as functions of the friction Reynolds number $Re_\tau$ can be generated. For the present test cases, such data are available from DNS in \cite{aiken2022}, which provides an opportunity to assess the predictive capabilities of the inverse framework beyond the original calibration point.

\begin{figure}[htbp]
  \centering
  \begin{subfigure}[b]{0.48\textwidth}
    \centering
    \includegraphics[width=1.00\textwidth,trim={1.2cm 6cm 1.5cm 6.5cm},clip]{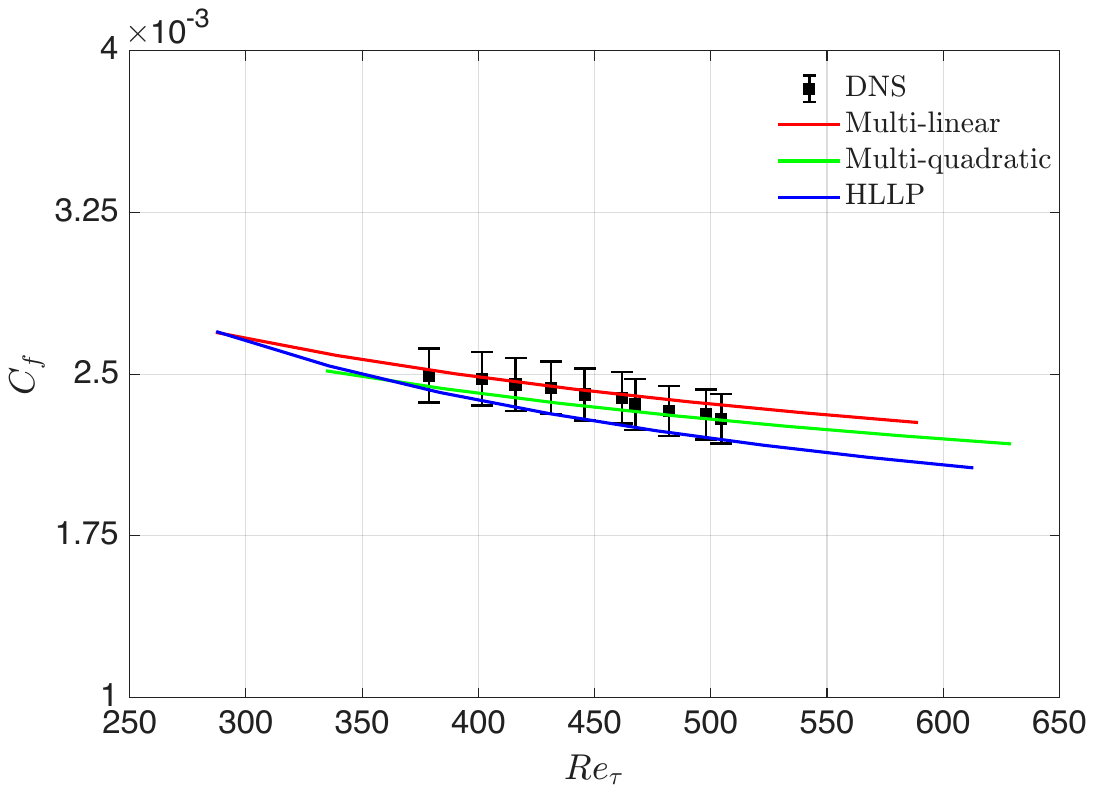}
    \caption{M2p5}
  \end{subfigure}
  \hfill
  \begin{subfigure}[b]{0.48\textwidth}
    \centering
    \includegraphics[width=1.00\textwidth,trim={1.2cm 6cm 1.5cm 6.5cm},clip]{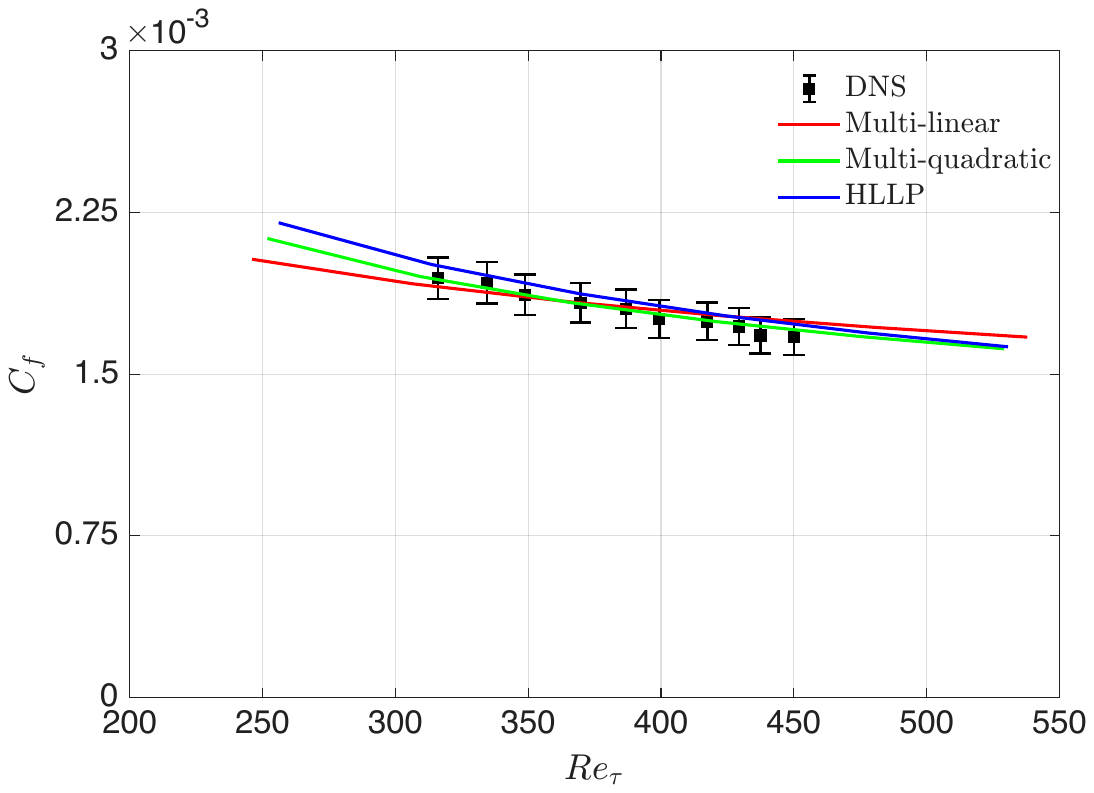}
    \caption{M6Tw025}
  \end{subfigure}
  \begin{subfigure}[b]{0.48\textwidth}
    \centering
    \includegraphics[width=1.00\textwidth,trim={1.2cm 6cm 1.5cm 6.5cm},clip]{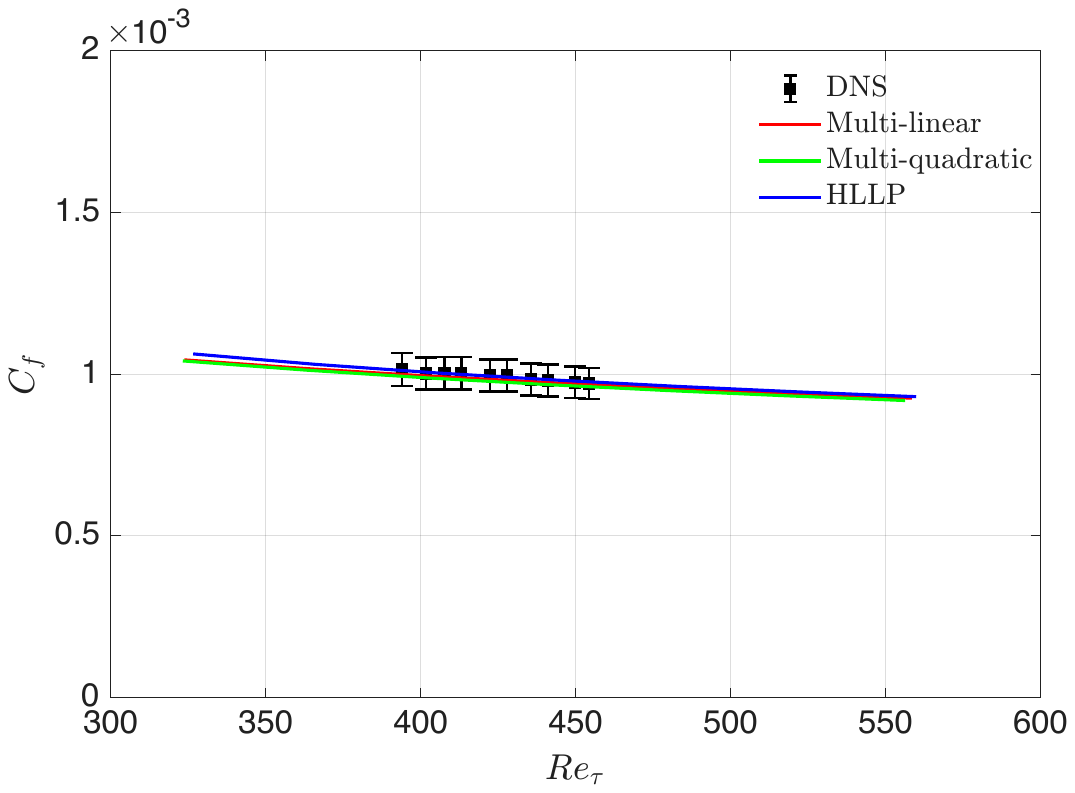}
    \caption{M6Tw076}
  \end{subfigure}
  \hfill
  \begin{subfigure}[b]{0.48\textwidth}
    \centering
    \includegraphics[width=1.00\textwidth,trim={1.2cm 6cm 1.5cm 6.5cm},clip]{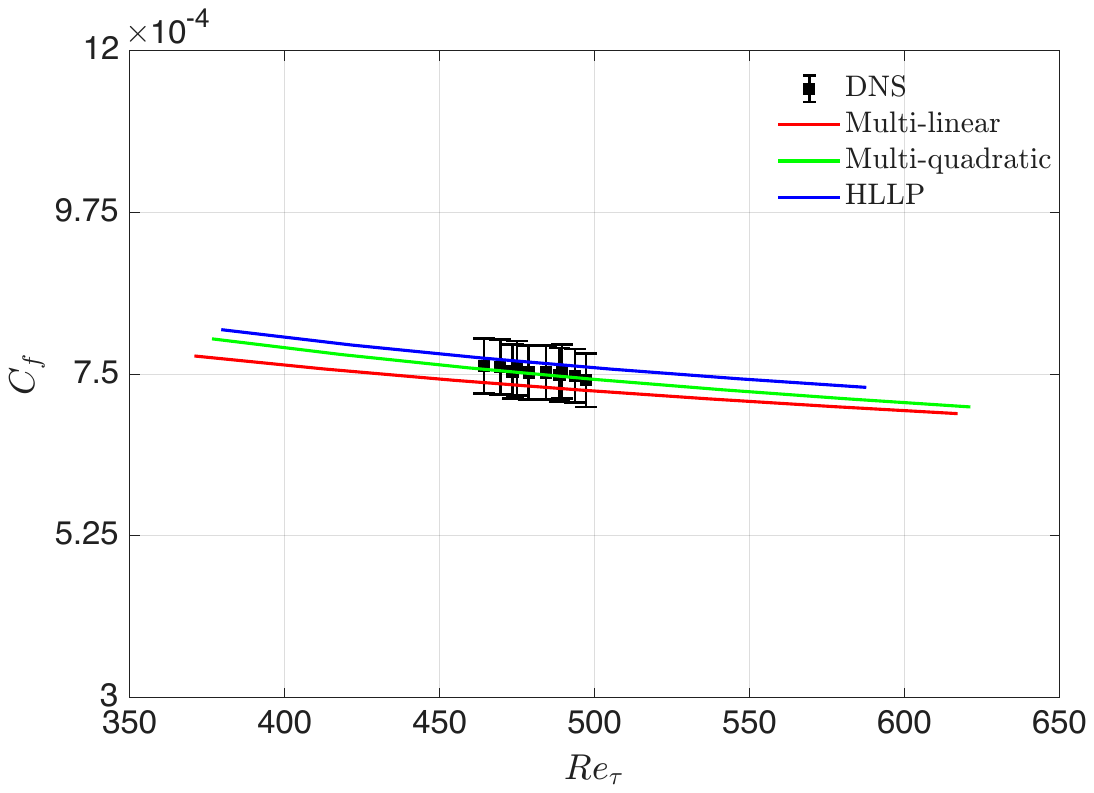}
    \caption{M8Tw048}
  \end{subfigure}
  \begin{subfigure}[b]{0.48\textwidth}
    \centering
    \includegraphics[width=1.00\textwidth,trim={1.2cm 6cm 1.5cm 6.5cm},clip]{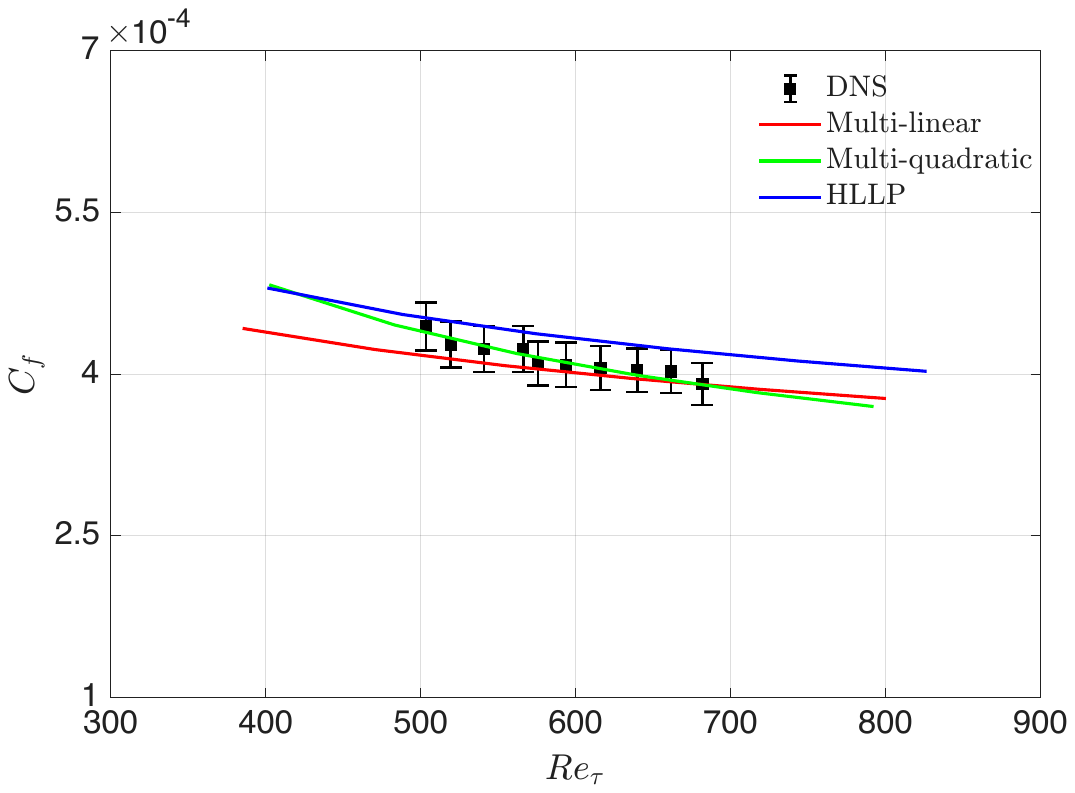}
    \caption{M14Tw018}
  \end{subfigure}
  \caption{Skin friction coefficient $C_f$ vs $Re_\tau$. The DNS data \cite{aiken2022} are shown with $\pm5\%$ error bars.}
  \label{fig:inverse_Cf_all}
\end{figure}

\begin{figure}[htbp]
  \centering
  \begin{subfigure}[b]{0.48\textwidth}
    \centering
    \includegraphics[width=1.00\textwidth,trim={1.2cm 6cm 1.5cm 6.5cm},clip]{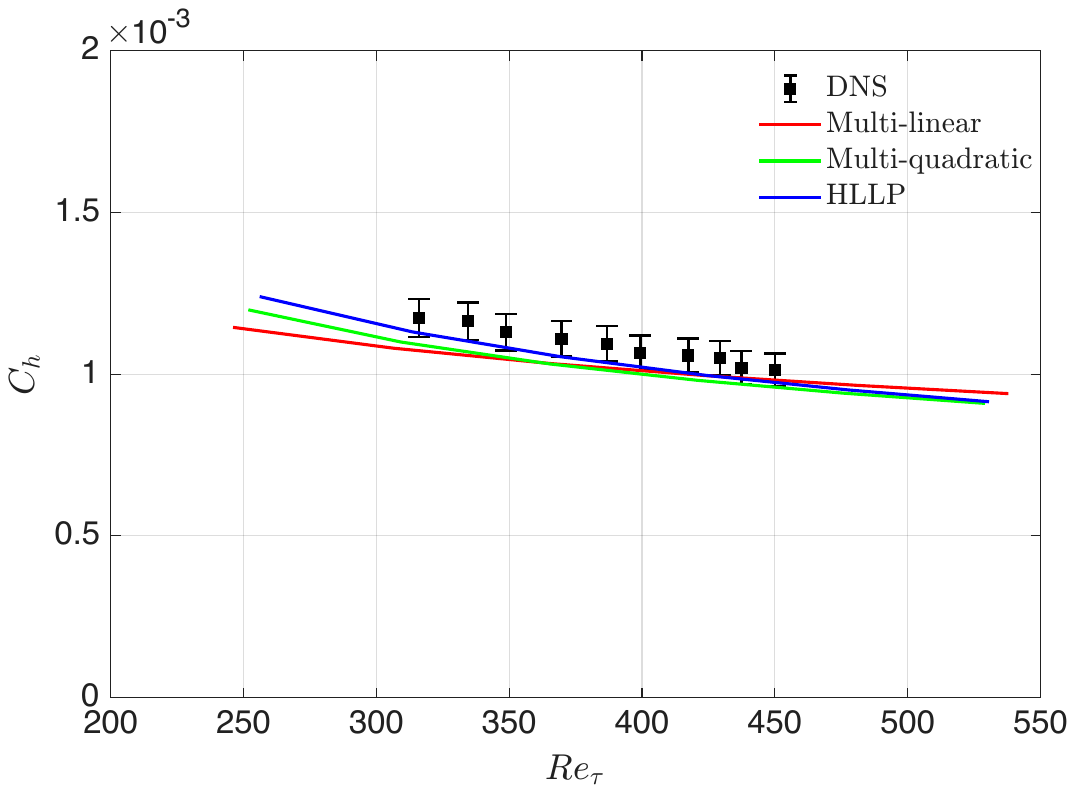}
    \caption{M6Tw025}
  \end{subfigure}
  \hfill
  \begin{subfigure}[b]{0.48\textwidth}
    \centering
    \includegraphics[width=1.00\textwidth,trim={1.2cm 6cm 1.5cm 6.5cm},clip]{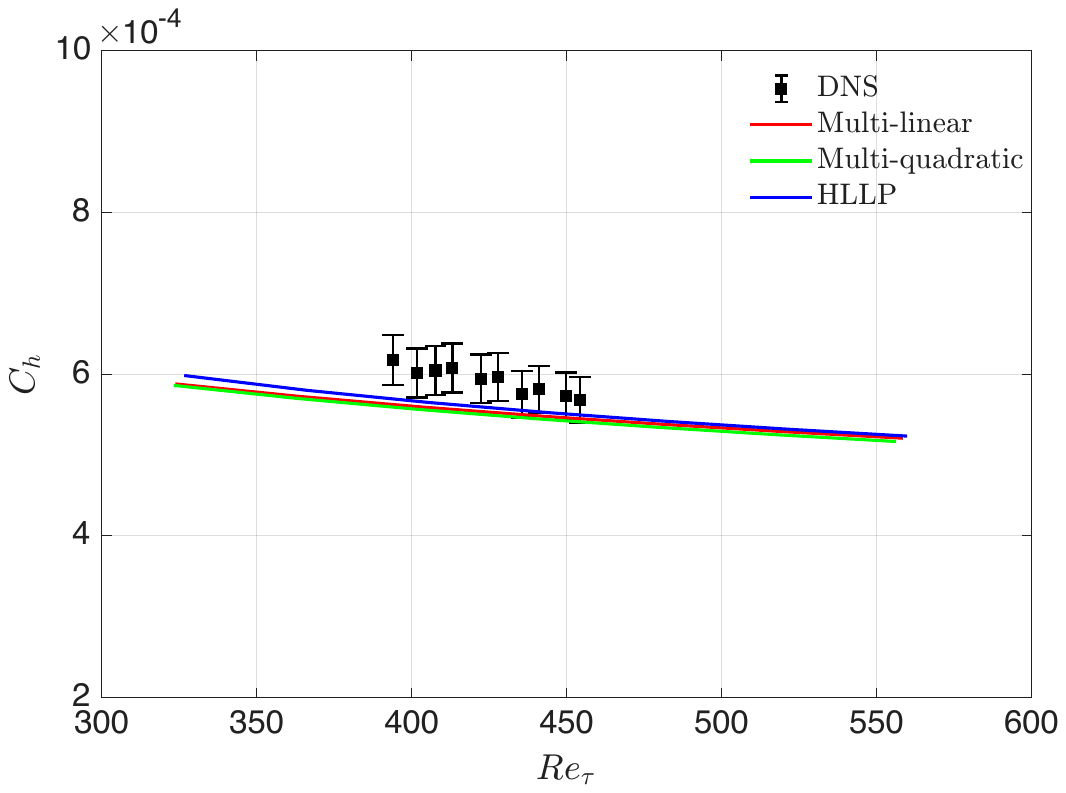}
    \caption{M6Tw076}
  \end{subfigure}
  \begin{subfigure}[b]{0.48\textwidth}
    \centering
    \includegraphics[width=1.00\textwidth,trim={1.2cm 6cm 1.5cm 6.5cm},clip]{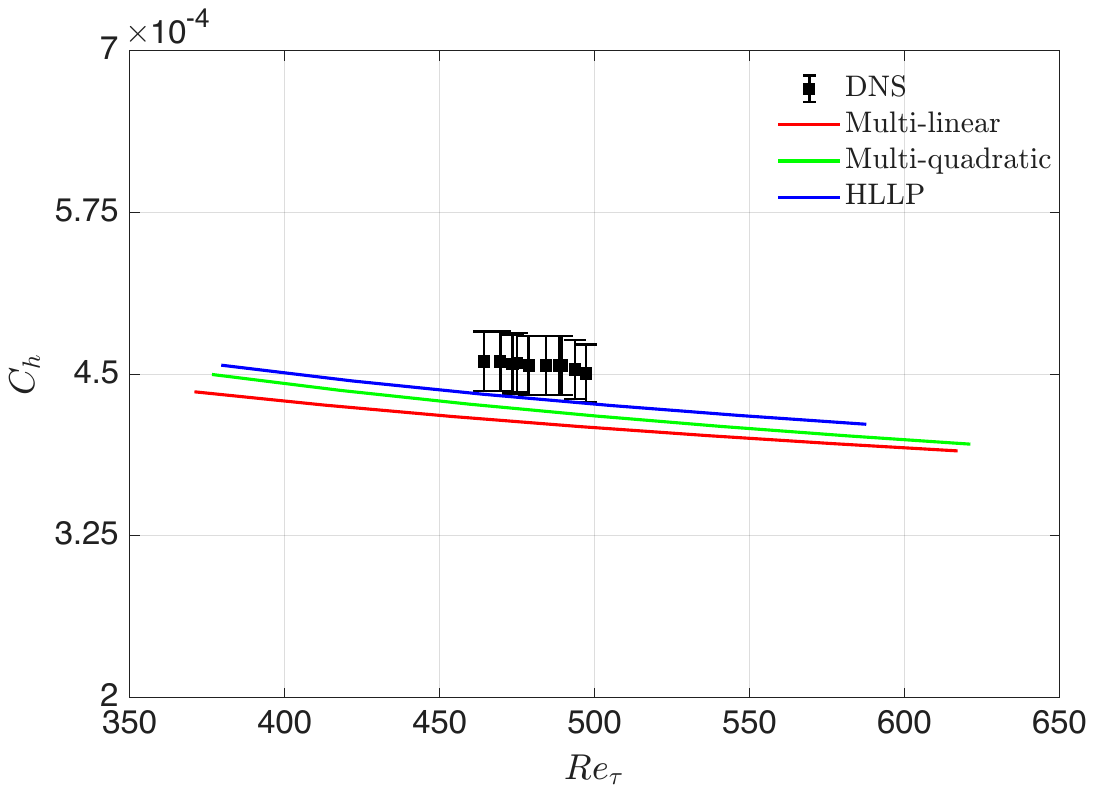}
    \caption{M8Tw048}
  \end{subfigure}
  \hfill
  \begin{subfigure}[b]{0.48\textwidth}
    \centering
    \includegraphics[width=1.00\textwidth,trim={1.2cm 6cm 1.5cm 6.5cm},clip]{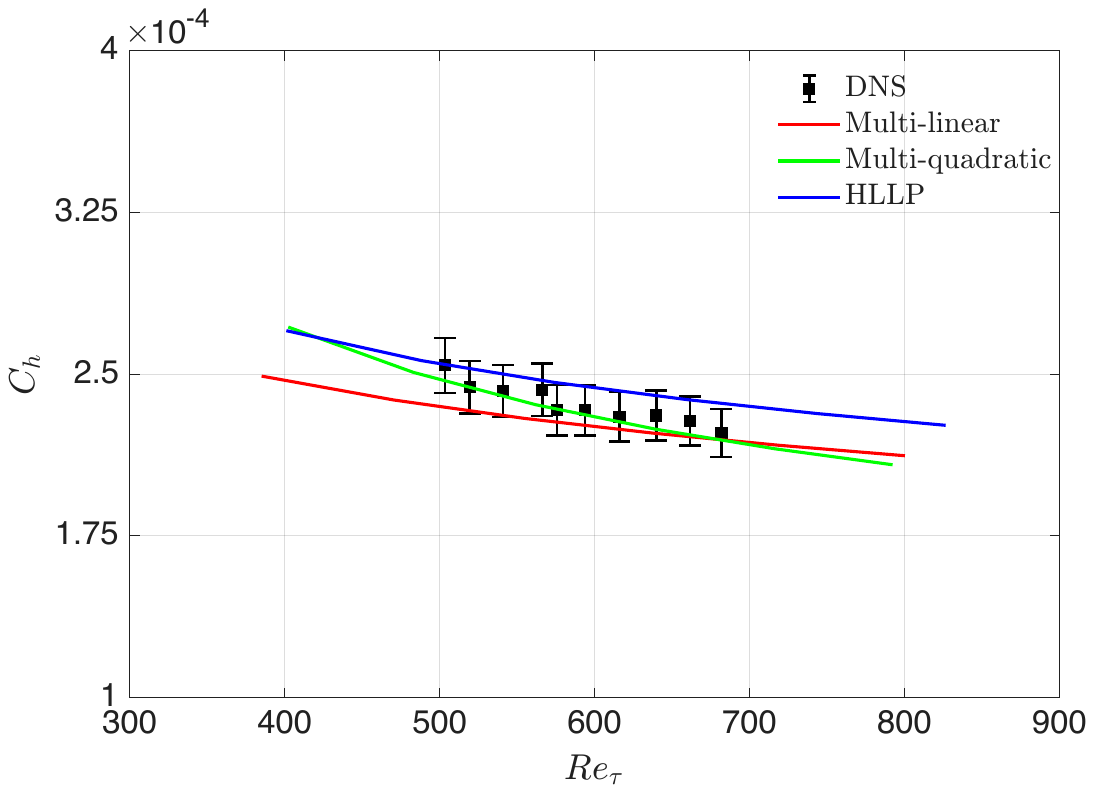}
    \caption{M14Tw018}
  \end{subfigure}

  \caption{Heat transfer coefficients $C_h$ vs $Re_\tau$. The DNS data \cite{aiken2022} are shown with $\pm5\%$ error bars.}
  \label{fig:inverse_Ch_all}
\end{figure}

This comparison is shown in \Cref{fig:inverse_Cf_all,fig:inverse_Ch_all}. The multi-quadratic model shows excellent performance for $C_f$, closely following $C_{f,\text{DNS}}$ with only minor discrepancies. The multi-linear model exhibits somewhat reduced accuracy, but its predictions remain within a $\pm 5\%$ error margin at all available $Re_\tau$ values. While both inverse models perform very well for $C_f$, their accuracy for $C_h$ is more modest, with the notable exception of case M14Tw018, for which the multi-quadratic model still achieves excellent agreement with DNS. The deterioration in $C_h$ predictions may be attributed to two factors: (i) the forward optimization targets only the mismatch between modeled and transformed velocity profiles and does not include any penalty term for temperature errors, and (ii) both the forward and inverse transformations rely on the TV relation in \Cref{eq:zhang-temperature}, rather than the actual DNS temperature profiles. Hence, improvements in the TV relation could potentially enhance $C_h$ predictions, but such refinements are beyond the scope of the present study.

\section{Conclusion}\label{sec:conclusion}

A framework has been developed for forward and inverse compressibility transformations of zero-pressure-gradient hypersonic turbulent boundary layers based on a stricter notion of consistency with an incompressible inner--outer model. The key requirement is that the transformed compressible mean velocity profile match a fixed incompressible inner--outer representation when expressed in terms of a transformed wall-normal coordinate, which in turn enforces collapse of semilocal eddy viscosity and turbulent kinetic energy production. Using this criterion, existing compressibility transformations (Griffin--Fu--Moin, Volpiani, and HLLP) were shown, for a set of hypersonic DNS cases, to exhibit maximum velocity errors ranging from a few percent up to about $25\%$ with respect to the adopted incompressible model, particularly in strongly cooled boundary layers. To address this limitation, a new forward transformation was introduced in which the transformed coordinate is formed as a convex combination of semilocal and integral-type basis functions, with coefficients correlated to the wall-based parameters $(M_\tau,B_q)$. Casewise optimization and regression showed that the proposed forward transformation achieves consistency errors of $1$--$4\%$ across the hypersonic database, while recovering expected trends such as increasing inner layer dominance and wake attenuation in the high Mach number, cold-wall limit.

The calibrated forward transformation was then embedded in an inverse solver that reconstructs compressible mean fields from freestream and wall conditions, a viscosity law, and a temperature--velocity relation. Formulated as a nonlinear problem for the wall shear stress and solved using a Newton iteration with complex step differentiation, the inverse framework recovers friction Reynolds numbers, mean velocity profiles, and skin friction coefficients in good agreement with DNS and generally improves upon the inverse HLLP model, especially for cold-wall cases. Predictions of heat transfer coefficients are more sensitive to the underlying temperature--velocity relation, but remain within reasonable agreement for the range of conditions considered. Overall, the results indicate that enforcing shear-level consistency in the transformed coordinate yields a more accurate and physically constrained mapping between compressible and incompressible states. Future work will extend the calibration database in Mach number, wall-to-recovery temperature ratio, and Reynolds number for improved heat transfer predictions, and assess the applicability of the proposed framework to  wall-modeled large eddy simulations of high-speed turbulent flows.

\clearpage
\bibliography{references}
\end{document}